\documentclass [journal,onecolumn,11pt]{IEEEtran}
\usepackage{amsfonts,amsmath,amssymb}
\usepackage{indentfirst, setspace}
\usepackage{url,float}
\usepackage{color}
\usepackage{xcolor}
\usepackage[a4paper,ignoreall]{geometry}
\usepackage{longtable}
\usepackage{graphicx}
\usepackage[a4paper,ignoreall]{geometry}
\usepackage{multicol}
\usepackage{stfloats}
\usepackage{enumerate}
\usepackage{cite}
\usepackage{amsthm}
\usepackage{multirow}
\usepackage{amssymb}
\usepackage{galois}
\usepackage[all]{xy}
\usepackage[square, comma, sort&compress, numbers]{natbib}
\usepackage{url,doi}
\usepackage{natbib,hyperref,doi}
\hypersetup{hidelinks}
\usepackage{comment}
\usepackage{verbatim}
\usepackage{tikz}
\usetikzlibrary{arrows}
\usepackage[justification=centering]{caption}

\def\qu#1 {\fbox {\footnote {\ }}\ \footnotetext { From Qu: {\color{red}#1}}}
\def\kq#1 {\fbox {\footnote {\ }}\ \footnotetext { From Quan: {\color{blue}#1}}}
\def\tl#1 {\fbox {\footnote {\ }}\ \footnotetext { From Niu: {\color{blue}#1}}}
\def\wang#1 {\fbox {\footnote {\ }}\ \footnotetext { From Wang: {\color{purple}#1}}}

\newtheorem{Th}{Theorem}[section]
\newtheorem{Cor}[Th]{Corollary}
\newtheorem{Prop}[Th]{Proposition}

\newtheorem{Lem}[Th]{Lemma}
\newtheorem{Def}[Th]{Definition}
\newtheorem{Exa}[Th]{Example}

\newtheorem{Rem}[Th]{Remark}

\newcommand{\tr}{{\rm Tr}}

\newcommand{\gf}{{\mathbb F}}

\makeatletter
\newcommand{\figcaption}{\def\@captype{figure}\caption}
\newcommand{\tabcaption}{\def\@captype{table}\caption}
\makeatother

\begin{document}

\title{Characterizations and constructions of $n$-to-$1$ mappings over finite fields}

		\author{
			{Tailin Niu, Kangquan Li, Longjiang Qu and Chao Li}
			\thanks{
				This work is supported 
				in part by the National Natural Science Foundation of China (NSFC) under Grant 62032009, Grant 62172427, 
				in part by the State Key Development Program for Basic Research of China under Grant 2019-JCJQ-ZD-351-00,
				in part by the Training Program for Excellent Young Innovators of Changsha under Grant kq1905052,
				in part by the Natural Science Foundation of Hunan Province of China under Grant 2021JJ40701,
				and in part by the Research Fund of National University of Defense Technology under Grant ZK20-42.
				\emph{(Corresponding author: Longjiang Qu.)}
				
				The authors are with the College of Liberal Arts and Sciences,
				National University of Defense Technology, Changsha, 410073, China (e-mail: runningniu@outlook.com; likangquan11@nudt.edu.cn; ljqu\_happy@hotmail.com; lichao\_nudt@sina.com).
				They are also with Hunan Engineering Research Center of Commercial Cryptography Theory and Technology Innovation, Changsha 410073, China.

%
			}
		}

	\maketitle{}
	\begin{abstract}		
$n$-to-$1$ mappings have wide applications in many areas, especially in cryptography, finite geometry, coding theory and combinatorial design.
In this paper, many classes of $n$-to-$1$ mappings over finite fields are studied.
First, we provide a characterization of general $n$-to-$1$ mappings over $\gf_{p^m}$ by means of the Walsh transform.
Then, we completely determine $3$-to-$1$ polynomials with degree no more than $4$ over $\mathbb{F}_{p^{m}}$.
Furthermore, we obtain an AGW-like criterion for characterizing an equivalent relationship between the $n$-to-$1$ property of a mapping over finite set $A$  and that of another mapping over a subset of $A$. 
Finally, we apply the AGW-like criterion into several forms of polynomials and obtain some explicit $n$-to-$1$ mappings.
Especially, three explicit constructions of the form $x^rh\left( x^s   \right) $ from the cyclotomic perspective, and several classes of $n$-to-$1$ mappings of the form $ g\left(  x^{q^k} -x +\delta   \right) +cx$ are provided.

	\end{abstract}

	
	\section{Introduction}
	Let $f$ be a mapping from one finite set $A$ to another finite set $ B$.
	Let $n$ be a positive integer.
	$f$ is called \textit{$n$-to-$1$} if for almost all $b \in B$, it has either $n$ or $0$ preimages of $A$ and for at most one exception $b_0 \in B$ (if exists), it has $t$ preimages with $0< t < n$.
	Let  $q$ be a prime power, and $\gf_q$ be the finite field with $q$ elements, and $ \gf_q^* $ denotes its all nonzero elements.
	$n$-to-$1$	mappings over finite fields ($ A=B=\gf_q $) have wide applications.
	When $n=1$, $n$-to-$1$ mappings become permutation polynomials (PPs).
	PPs and their compositional inverses, as well as their applications over finite fields have been extensively studied.
	For some literature on permutation polynomials over finite fields, we refer to \cite{hou2015permutation,wang201915,li2019survey} and the references therein.
	When $n\ne 1$, $n$-to-$1$ mappings also have useful applications in large areas, especially in cryptography, finite geometry, coding theory and combinatorial design.
	For	example, the well known bent functions, class $\mathcal{H}$ can be constructed from  $2$-to-$1$ mappings over finite fields with characteristic $ 2 $, by Carlet and Mesnager \cite{carlet2011dillons,mesnager2016class}. 
	Univariate Niho bent functions were constructed from the well-known o-polynomial, a class of polynomial characterized by $2$-to-$1$ mappings \cite{budaghyanUnivariateNihoBent2016}.
	Despite constructing bent functions \cite{mesnager2015bent,budaghyan2014niho} and semi-bent functions \cite{mesnager2013semi},
	o-polynomials also have relations with some spreads of presemifields \cite{cesmelioglu2015bent}.
	In \cite[Section 6]{mesnagerTwotoOneMappingsFinite2019a}, Mesnager and Qu proposed some bent functions, semi-bent functions and planar functions by some $2$-to-$1$ mappings.
	In addition, $2$-to-$1$ mappings can be used to construct difference sets \cite{ding2014codes} and binary linear codes \cite{ding2016construction,mesnager2015bent,liBinaryLinearCodes2021}.
	Furthermore, in \cite{dalai3to1PowerAPN2008} Dalai proposed several $3$-to-$1$ APN S-boxes.
	Many known APN functions are also founded being $3$-to-$1$ in \cite{dalai3to1PowerAPN2008,budaghyan2021triplicate}.
	In 2021, Budaghyan et al. \cite{budaghyan2021triplicate} used triplicate functions, a generalization of $3$-to-$1$ mappings, to study APN functions.


    Due to the importance of $n$-to-$1$ mappings, some scholars' research interests were attracted.
	Mesnager and Qu \cite{mesnagerTwotoOneMappingsFinite2019a} provided a detailed and systematic study of $2$-to-$1$ mappings over finite fields.
	They characterized  $2$-to-$1$ mappings by the Walsh transform and determined $2$-to-$1$ polynomials in $\gf_q$ with degree no more than $4$.
	Many explicit constructions of $2$-to-$1$ mappings were also given.
	In \cite{liFurtherStudy2to12021}, Li et al. pushed further the study of $2$-to-$1$ mapping initiated in \cite{mesnagerTwotoOneMappingsFinite2019a}.
	They focused on binary fields, and determined $2$-to-$1$ polynomials with degree $5$ using the Hasse-Weil bound.
	Besides, mainly $2$-to-$1$ trinomials and quadrinomials over $\gf_{2^m}$ were obtained.
	Recently, Gao et al. \cite{gaoMto1MappingsFinite2021} generalized some works in \cite{mesnagerTwotoOneMappingsFinite2019a,liFurtherStudy2to12021} to $n$-to-$1$ mapping over finite set $A$ with $\# A \equiv 0  \pmod n  \text { or } \# A \equiv 1 \pmod n$.
	They mainly considered such $n$-to-$1$ mappings from PPs and obtained some constructions.
	Yuan et al. \cite{yuanTwotooneMappingsInvolutions2021b} considered to obtain $2$-to-$1$ mappings from other ones over subsets.
	Several explicit $2$-to-$1$ mappings of the form $ g\left(  x^{q} -x +\delta   \right) +x$ over $\gf_{2^m}$ were also constructed by them.
	They also used these constructions to obtain involutions over $\gf_{2^m}$.
	However, there are few characterization of general $n$-to-$1$ mappings or equivalent $n$-to-$1$ properties between them in the literature.
	This motivates us to generalize the work above and provide a unified study of $n$-to-$1$ mappings over finite fields.

    The main purpose of this paper is to provide approaches to constructing several types of explicit $n$-to-$1$ mappings over finite fields.
	First, we provide a general definition of $n$-to-$1$ mappings and a characterization of general  $n$-to-$1$ mappings over $\gf_{p^m}$ by means of the Walsh transform.
	This characterization utilizes the properties of a special polynomial to represent $n$-to-$1$ mappings.
	Then, we completely determine $3$-to-$1$ mappings with degree no more than $4$ over $\mathbb{F}_{p^{m}}$.
	Low degree $3$-to-$1$ polynomials will be helpful as basic results for constructing AGW-like $n$-to-$1$ mappings.
	After that, inspired by the AGW-like criterion, we find that there is an equivalent relationship between $n$-to-$1$ mappings over different sets for general $n$.
 	Thus, an AGW-like criterion for constructing $n$-to-$1$ mappings is obtained, which can be viewed as a kind of generalization of the AGW criterion and the work in \cite[Proposition 4.2]{yuanTwotooneMappingsInvolutions2021b}. 	
	Then, we apply Theorem \ref{agwcore2} into different forms of polynomials, e.g., $ h(\psi(x)) \phi(x)+g(\psi(x))$, $L_{1}(x)+L_{2}(x) g(L_{3}(x))$, $x^rh(x^s)$ and $ g\left(  x^{q^k} -x +\delta   \right) +cx$, 
	and obtain several constructions of $n$-to-$1$ mappings.
	For the polynomials of the form $x^rh(x^s)$ from the cyclotomic perspective, we use a piecewise method to provide three explicit constructions.
	In addition, approaches to obtaining new $n$-to-$1$ mappings reductively from known $n$-to-$1$ mappings or permutations over their subfields are obtained.
	For polynomials of the form $g\left(  x^{q^k} -x +\delta   \right) +cx$, the necessary and sufficient conditions for five explicit constructions being $n$-to-$1$ are obtained.

The remainder of this article is organized as follows.
In Section \ref{walsh}, we provide a definition and a characterization of general $n$-to-$1$ mapping by means of the Walsh transform.
Section \ref{lowdegree}  determine $3$-to-$1$ polynomials with degree no more than $4$.
In Section \ref{agwlike}, we propose an AGW-like criterion for $n$-to-$1$ mappings and handle constructions of the form $ h(\psi(x)) \phi(x)+g(\psi(x))$, $L_{1}(x)+L_{2}(x) g(L_{3}(x))$, $x^rh(x^s)$ and $ g\left(  x^{q^k} -x +\delta   \right) +cx $.
Throughout this article, for any integer $ m \ge2 $, $\tr_{q^m/q}$ denotes the trace function from $\gf_{q^m}$ to $\gf_{q}$, i.e., $\tr_{q^m/q}(x)=\sum_{i=0}^{m-1} x^{q^i}$ for any $x \in \gf_{q^m}$.
We use $\#A$ and $A^*$ to denote the cardinality of a set $A$ and $ A \setminus \{0\} $ respectively.
For a real number $a$, $ \lceil a \rceil $ denotes the least integer greater than or equal it.

%

\section{ a characterization by means of the Walsh transform}
\label{walsh}

%
%

In this section, we present the definition of $n$-to-$1$ mappings, and a characterization by means of the Walsh transform.

In \cite{gaoMto1MappingsFinite2021}, Gao et al. defined $n$-to-$1$ mappings with $\# A \equiv 0  \pmod n  \text { or } \# A \equiv 1 \pmod n$.
Here we provide a definition without the limitation in $\#A$.
\begin{Def}
Let $f$ be a mapping from one finite set $A$ to another finite set $B$.
Then $ f $ is called an $n$-to-$1$ mapping if one of the following two cases holds:
\begin{enumerate}[(1)]
	\item if $ n \mid \# A  $, and for any $ b \in B$, it has either $n$ or $ 0 $ preimages of $f$ in $A$.
	\item if $ n \nmid \# A$, and for almost all $b \in B$,  it has either $n$ or $ 0 $ preimages of $f$ in $A$, and for the only one exception element, it has exactly
	$ \# A \mod  n $ preimages.
\end{enumerate}
\end{Def}

In the following, we discuss  $n$-to-$1$ mappings defined over finite fields ($ A=B= \gf_{p^m} $).
Below we present some basic results.
\begin{Lem}\label{monomial}
	Let $f(x)=a x^{d}$ be a monomial polynomial over $\mathbb{F}_{q}$, where $a \neq 0$.
	Then $f$ is $n$-to-$1$ over $\gf_{q}$ if and only if $\gcd(d, q-1)=n$.
\end{Lem}

The lemma below is a direct conclusion from \cite[Theorem 3.50]{lidl1997finite}.
\begin{Lem}\label{linear}
	A $q$-polynomial $L(x) $ is $q^{m-k}$-to-$1$ over $\gf_{q^m}$, where $k$ is the rank of $L$ as a linear transform.
\end{Lem}

The $n$-to-$1$ property of Dickson polynomials has been studied in literature.
Interest readers can refer to \cite{lidlDicksonPolynomials1993,hou2009reversed,dillon2004new}.

After that, we consider a characterization by means of the Walsh transform.
Let  $F : \gf_{p^m} \rightarrow \gf_{p^m}  $  be an $n$-to-$1$ mapping, and $\omega$ be a $p$-th primitive root of $1$, where $p$ is a prime and $ m $ is a positive integer.
The Walsh transform of $F$ at $(u, v) \in \mathbb{F}_{p^{m}} \times \mathbb{F}_{p^{m}}$ equals by definition the Walsh transform of the so-called component function $\tr_{p^{n} / p}(v F(x))$ at $u$, that is:
$$
W_{F}(u, v):=\sum_{x \in \mathbb{F}_{p^{m}}}\omega^{\tr_{p^{m}/ p} (v F(x))+\tr_{p^{m}/ p}(u x) }.
$$
There are some differences between situations $ n \mid  p^m  $ and $ n \nmid  p^m$ for $F$ being $n$-to-$1$, when we characterize $F$ by the Walsh transform.

We first consider the situation when $ n \nmid  p^m$.
Let $\phi(X)=\sum_{j \geq 0} A_{j} X^{j}$ be any polynomial over $\mathbb{R}$ such that
\begin{equation}\label{gouzaophi1}
	\phi(X)
	 \begin{cases}
		=0, & \text { if } X=0, n \\
		=1, & \text { if } X=p^m \mod n \\
		>1, & \text { if } X \in \mathbb{N} \backslash\{0, \ p^m \mod n, \ n\} .\\
	\end{cases}
\end{equation}
 Hence for any $F$ and $b \in \mathbb{F}_{p^{m}}$, we have
$$
\sum_{j \geq 0} A_{j}\left(\#\left\{x \in \mathbb{F}_{p^m}: F(x)-b=0\right\}\right)^{j} \geq 0
$$
and $F$ is an $n$-to-$1$ function if and only if this inequality is an equality for almost all $b \in \mathbb{F}_{p^{m}}$ with exactly one exception.
Furthermore, for any $F$, we have
$$\sum_{j \geq 0} A_{j} \sum_{b \in \mathbb{F}_{p^m}}\left(\#\left\{x \in \mathbb{F}_{p^m}: F(x) - b=0\right\}\right)^{j} \geq 1$$
and $F$ is $n$-to-$1$ if and only if this inequality is an equality.
We shall now characterize this condition by means of the Walsh transform.
It is clear that
\begin{equation*}
	\begin{aligned}
		&\#\left\{x \in \mathbb{F}_{p^m}: F(x)-b=0\right\} 	=  p^{-m} \sum_{x \in \mathbb{F}_{p^m}, v \in \mathbb{F}_{p^m}}    \omega^{\tr_{p^m/ p}(v(F(x)   -  b))}
	\end{aligned}
\end{equation*}
and therefore for $j  \ge1$, we have
\begin{equation*}
	\begin{aligned}
		&     \sum_{b \in \mathbb{F}_{p^m}}  \left(\#\left\{x \in \mathbb{F}_{p^m}: F(x) - b=0\right\}\right)^{j}      \\
		=&  p^{-jm}           \sum_{b \in \mathbb{F}_{p^m}}                       \sum_{	
			x_1, \cdots,x_j \in \mathbb{F}_{p^m},  \atop	v_{1}, \cdots, v_{j} \in \mathbb{F}_{p^m}
		}
		\omega^{    \sum_{i=1}^{j}       \tr_{p^m/ p}(v_i(F(x_i)   -  b))}       \\
		=&  p^{-jm}                       \sum_{	
			x_1, \cdots,x_j \in \mathbb{F}_{p^m},  \atop		v_{1}, \cdots, v_{j} \in \mathbb{F}_{p^m}
		}
		\left(
		\omega^{    \sum_{i=1}^{j}       \tr_{p^m/ p}(v_iF(x_i)  )}
		\cdot \sum_{b \in \mathbb{F}_{p^m}}
		\omega^{    \sum_{i=1}^{j}       \tr_{p^m/ p}((  -  b)v_i)}
		\right)   \\	
		=&  p^{m-jm}                       \sum_{	
			v_{1}, \cdots, v_{j} \in \mathbb{F}_{p^m}     \atop		\sum_{i=1}^{j} v_{i}=0
		}
		\prod_{i=1}^{j} W_{F}(0, v_{i}) . \\	
	\end{aligned}
\end{equation*}
Hence we have the following characterization of $n$-to-$1$ mappings over $\gf_{p^m}$ by the Walsh transform.

\begin{Th}
	Let $F: \mathbb{F}_{p^m} \rightarrow \mathbb{F}_{p^m}$ be an $n$-to-$1$ mapping, where $ n \nmid  p^m$.
	Then
	$$
	A_{0}+\sum_{j \geq 1} A_{j} p^{m-jm} \sum_{v_{1}, \cdots, v_{j} \in \mathbb{F}_{p^m} \atop \sum_{i=1}^{j} v_{i}=0} \prod_{i=1}^{j} W_{F}\left(0, v_{i}\right) \geq 1
	$$
	and F is $n$-to-$1$ if and only if this inequality is an equality, where $A_j$ are coefficients of $\phi(X)=\sum_{j \geq 0} A_{j} X^{j}$ satisfying the condition  (\ref{gouzaophi1}).
\end{Th}
It is not hard to find a polynomial satisfying $\phi$ in the condition (\ref{gouzaophi1}), and below is an example.
For $2$-to-$1$ mappings over $\gf_{p^m}$ with $p$ is odd, consider the polynomial $\phi_1(X)= X(X-2)^2 =X^3-4 X^2+  4 X$ over $ \mathbb{N} $.
It takes value $1$ when $X=1 $ and $\phi_1(X) >1 $ when $ X \in \mathbb{N}  \backslash\{0,1,2\}  $, where $\mathbb{N}$ denotes the set of all no-negative integers.
Thus we have the following corollary.
 \begin{Cor}
 	Let $F: \mathbb{F}_{p^m} \rightarrow \mathbb{F}_{p^m}$ be an $n$-to-$1$ mapping, where $p$ is odd.
 	Then
 	\begin{equation*}\begin{aligned}
 			&p^{-2m} \sum_{v_{1}, v_2 \in \mathbb{F}_{p^m}} W_{F}\left(0, v_{1}\right)W_{F}\left(0, v_{2}\right)W_{F}\left(0, -v_1-v_2\right)     \\
 			&-4 p^{-m} \sum_{v \in \mathbb{F}_{p^m}  }   W_{F}\left(0, v\right)W_{F}\left(0, -v\right)   + 4  p^m        \geq 1
 	\end{aligned}\end{equation*}
 	and F is $2$-to-$1$ if and only if  this inequality is an equality.
 \end{Cor}

When $ n \mid p^m$, i.e., $n=p^k$ with $1\le k \le m$, we only need the polynomial $\phi(X)=\sum_{j \geq 0} A_{j} X^{j}$ satisfying
\begin{equation}\label{gouzaophi2}
	\phi(X)
	\begin{cases}
		=0, & \text { if } X=0, p^k \\
		>0, & \text { if } X \in \mathbb{N} \backslash\{0,  \ p^k\} .\\
	\end{cases}
\end{equation}
By a similar deduction, we can obtain
$$\sum_{j \geq 0} A_{j} \sum_{b \in \mathbb{F}_{p^m}}\left(\#\left\{x \in \mathbb{F}_{p^m}: F(x) - b=0\right\}\right)^{j} \geq 0$$
and the following theorem.
\begin{Th}
	Let $F: \mathbb{F}_{p^m} \rightarrow \mathbb{F}_{p^m}$ be an $n$-to-$1$ mapping.
	Then
	$$
	A_{0}+\sum_{j \geq 1} A_{j} p^{m-jm} \sum_{v_{1}, \cdots, v_{j} \in \mathbb{F}_{p^m} \atop \sum_{i=1}^{j} v_{i}=0} \prod_{i=1}^{j} W_{F}\left(0, v_{i}\right) \geq 0
	$$
	and F is $p^k$-to-$1$ if and only if this inequality is an equality, where $1\le k \le m$ and $A_j$ are coefficients of $\phi(X)=\sum_{j \geq 0} A_{j} X^{j}$ satisfying the condition  (\ref{gouzaophi2}).
\end{Th}
We consider the polynomial $\phi_2(X) = X(X-p^k)^2 = X^3-2 p^k X^2+  p^{2k}  X $ over $ \mathbb{N} $.
Clearly, it takes value $0$ when $X$ equals $ 0 $ or $ p^k $ and takes strictly positive value when $ X $ is in $\mathbb{N}  \backslash\{0,2\}  $.
Then, we have the following corollary.
\begin{Cor}
	Let $F: \mathbb{F}_{p^m} \rightarrow \mathbb{F}_{p^m}$ be an $n$-to-$1$ mapping.
	Then
	\begin{equation*}\begin{aligned}
			&p^{-2m} \sum_{v_{1}, v_2 \in \mathbb{F}_{p^m}} W_{F}\left(0, v_{1}\right)W_{F}\left(0, v_{2}\right)W_{F}\left(0, -v_1-v_2\right)     \\
			&-2n p^{-m} \sum_{v \in \mathbb{F}_{p^m}  }   W_{F}\left(0, v\right)W_{F}\left(0, -v\right)   + p^{2k}      p^m        \geq 0
	\end{aligned}\end{equation*}
	and F is $p^k$-to-$1$ if and only if this inequality is an equality, where $1\le k \le m$.
\end{Cor}

\section{low degree $3$-to-$1$ polynomials}
\label{lowdegree}

In this section, we completely determine  $3$-to-$1$ mappings with degree no more than $4$ over $\mathbb{F}_{p^{m}}$.
Clearly, for any polynomial $f(x) \in \mathbb{F}_{p^{m}}[x]$ with degree $d$, it is $n$-to-$1$ over $\mathbb{F}_{p^{n}}$ if and only if so is $g(x)=a f(x+b)+c$, where $a \in \mathbb{F}_{p^{n}}^{*}$ and $b, c \in \mathbb{F}_{p^{m}}$.
Thus, it is sufficient to consider $f(x) \in \mathbb{F}_{p^{n}}[x]$ with the normalized form, i.e., $f(x)$ is monic, $f(0)=0$, and when $\gcd(p, d)=1$, the coefficient of $x^{d-1}$ is $0$.

\begin{Th}
	\label{de3p3}
	$f(x)=x^{3}+ax^2+bx $ over $\mathbb{F}_{3^{m}}$ is $3$-to-$1$ if and only if
	$a=0$ and $-b$ is a square in $\gf_{3^m}^*$, where $a, b \in \mathbb{F}_{p^{m}}$
\end{Th}
\begin{proof}
	It suffices to prove for any $x \in \gf_{3^m}$,	the equation
	$	f(x+y)=f(x)$
	has exactly three distinct solutions in $\gf_{3^m}$ for the variable $y$.
	Expanding $	f(x+y)=f(x)$, we obtain
	$$x^{3}+y^{3}+a   x^2+ay^2-axy      +by+bx=x^{3}+ax^2+bx.$$
	After simplifying it, one can obtain
	$$y^{3} +ay^2-axy      +by=0.$$
	Clearly $y=0$ is a solution, and we assume $y\ne 0$ below. Then we have
	$$y^{2} +ay   -ax      +b=0,$$
	which is equivalent to
	\begin{equation*}\label{fasdad}
		(y-a)^2=a^2+ ax  - b.
	\end{equation*}
	It is clear that the above equation has two distinct solution for any $x \in \gf_{p^m}$ if and only if $a=0$ and $-b$ is a square in $\gf_{3^m}^*$.
\end{proof}

Before we consider $p\ne 3$, we firstly recall a lemma.
\begin{Lem}\cite{lidl1997finite}\label{2gen}
	Let $a, b \in \mathbb{F}_{2^{n}}$ and $a \neq 0$. 
	Then the quadratic equation $x^{2}+a x+b=0$ has solutions in $\mathbb{F}_{2^{n}}$ if and only if $\tr_{2^n/2}\left(\frac{b}{a^{2}}\right)=0$.
\end{Lem}
When $p\ne 3$, it suffices to consider $f(x)=x^{3}+bx $, where $b \in \mathbb{F}_{p^{m}}$.
\begin{Th}
	\label{de3pne3}
	For $p \ne 3$, $f(x)=x^{3}+bx $ over $\mathbb{F}_{p^{m}}$ is $3$-to-$1$ if and only if $b=0$ and
	\begin{enumerate}[(1)]
		\item $p=2$, $m$ is even; or
		\item $p>3$, $3 \mid ( p^m-1)$.
	\end{enumerate}
\end{Th}
\begin{proof}
	It suffices to prove for almost all $x \in \gf_{p^m}$,	
	$	f(x+y)=f(x)$
	has exactly three distinct solutions for the variable $y$, with at most one exception $x_0$ in $\gf_{p^m}$ has one or two distinct solutions.
	Expanding and simplifying $f(x+y)=f(x)$, we obtain
	$$ 3x^2y+3xy^2 +y^{3}     +by=0$$
	Clearly $y=0$ is a solution, and we assume $y\ne 0$ below. Then we have
	\begin{equation}\label{adfasdfasdffwefsdfcsgafsd}
		y^{2}   +3xy +      3x^2     +b=0.
	\end{equation}
	If $p=2$ and $x=0$, Eq. (\ref{adfasdfasdffwefsdfcsgafsd}) is reduced to $y^2=b$, which has only one solution.
	If $p=2$ and $x \ne 0$, let $y=xz$.  
	Then Eq. (\ref{adfasdfasdffwefsdfcsgafsd}) is reduced to
	\begin{equation}\label{kavasdasdey}
		z^2+z+1+\frac{b}{x^2}=0.
	\end{equation}
	By lemma \ref{2gen}, Eq. (\ref{kavasdasdey}) have solutions in $\gf_{2^m}$ if and only if $\tr_{2^m/ 2}(1+\frac{b}{x^2})=0$.
	If $b \ne 0$, $\tr_{2^m/ 2}(1+\frac{b}{x^2})=1$ can not establish for any $x \in \gf_{2^m}^*$.
	Thus $b=0$ and $m$ is even.
	
	If $p > 3$, multiplying the both side of Eq. (\ref{adfasdfasdffwefsdfcsgafsd}) with $4$, we obtain
	$$  4y^2+   9x^2    + 12xy =-3x^2-4b ,   $$
	i.e., 
	$$    (2y+3x)^2=-3x^2-4b.    $$
	We assume for any $x\in \gf_{p^m}$ but one exception, $-3x^2-4b$ is a square.
	Then there exists a polynomial $P(x)$ over $\gf_{p^m}$ satisfying $P(x)^2=-3x^2-4b$, for almost all $x\in \gf_{p^m}$ (ignore one exception).
	If $b \ne 0$, it is impossible.
	Then we obtain $f(x)=x^3$ and $\gcd(3,p^m-1)=3$.
	Thus $f(x+y)=f(x)$ has three distinct solutions if and only if $b=0$, $3  \mid ( p^m-1)$.

		In summary, $f $ is $3$-to-$1$ if and only if $p>3$, $b=0$ and $-3$ is a square in $\gf_{p^m}$; or $p=2$, $m$ is even.
	\end{proof}

	\begin{Th}
		For $p^m > 15$,
		$f(x)=x^{4}+ax^3+bx^2+cx $ over $\mathbb{F}_{p^{m}}$ is not $3$-to-$1$.
	\end{Th}
	\begin{proof}
		Assume $f$ is $3$-to-$1$, then $f(x)-d=0$ have and only have $3$ distinct roots for all but at most one exception $d \in \gf_{p^m}$. 
		Since the degree of $f$ is $4$,	 the polynomial $f(x)-d$  must have multiple roots.
		Thus $$\gcd(f(x)-d , f'(x)) \ne 1  .$$
		That is,
		\begin{equation}\label{habsudhvauscvxyutc}
			\gcd(x^4+ax^3+bx^2+cx-d, 4x^3+3ax^2+2bx+c) \ne 1.
		\end{equation}
		
		If $p = 2$, then the condition (\ref{habsudhvauscvxyutc}) is reduced to
		\begin{equation}\label{baishdbxiauhs}
			\gcd(x^4+ax^3+bx^2+cx-d,  ax^2 +c) \ne 1.
		\end{equation}
		If $ac\ne0$, $f'(x)$ has at most $2$ distinct roots, and thus $d$ has at most $2$ values.
		Then, we have
		$$2 \ge \# \{ d  \  | d \text{ satisfies the condition (\ref{baishdbxiauhs})}  \   \}  \ge \lceil \frac{p^m}{3} \rceil   -1. $$
		Thus, we get $2^m \le 12$.
		If $a=c=0$, $f(x)=x^{4}+bx^2 $ is linear.
		By Lemma \ref{linear}, $f$ can not be $3$-to-$1$.

		If $p\ne 2$, $f'(x)$ has at most $3$ distinct roots, and thus $d$ has at most $3$ values by the condition (\ref{habsudhvauscvxyutc}).
		Then, we obtain
		$$3 \ge \# \{ d  \  | d \text{ satisfies the condition (\ref{habsudhvauscvxyutc})}  \   \}  \ge \lceil \frac{p^m}{3} \rceil -1  . $$
		Thus, we obtain $p^m \le 15$.
		In summary, $f(x)=x^{4}+ax^3+bx^2+cx $ is not $3$-to-$1$ over $\gf_{p^m}$, for any $p^m > 15$.
	\end{proof}
	\begin{Rem}
		By a Magma program searching, we find that there are no $3$-to-$1$ mappings of the form $f(x)=x^4+ax^3+bx^2+cx$ being $3$-to-$1$ when $ 4<p^m \le 15$.
\end{Rem}



\section{an AGW-like criterion for $n$-to-$1$ mappings and explicit constructions}
\label{agwlike}


The AGW criterion establishes an equivalent connection between the permutation property of a mapping over a finite set $A$ and the permutation property of another relate mapping over subsets of $A$.
\begin{Lem}
	\label{LGWlemma}
	(\cite{akbary2011constructing}, AGW Criterion)
	Let $A, S$, and $\overline{S}$ be finite sets with $\# S=\# \overline{S}$, and let $f: A\to A,$ $g: S\to \overline{S}$, $\lambda: A\to S$ and $\overline{\lambda}: A\to\overline{S}$ be mappings such that $\bar{\lambda}\circ f=g\circ \lambda$. If both $\lambda$ and $\bar{\lambda}$ are surjective, then the following statements are equivalent:
	\begin{enumerate}[(1)]
		\item $f$ is a bijection and
		\item $g$ is a bijection from $S$ to $\overline{S}$ and $f$ is injective over $\lambda^{-1}(s)$ for each $s\in S$.
	\end{enumerate}
\end{Lem}
It can be illustrated as the following commutative diagram:
\begin{equation*}
	\xymatrix{
		A \ar[rr]^{f}\ar[d]_{\lambda} &   &  A  \ar[d]^{\overline{\lambda}} \\
		S	 \ar[rr]^{g} &  & \overline{S} .}
\end{equation*}
Since the AGW criterion was put forward, it attracted much attention and many PPs based on it are constructed.
Recently, this technique has been generalized to construct $2$-to-$1$ mappings over $\gf_{2^n}$ by Mesnager and Qu \cite{mesnagerTwotoOneMappingsFinite2019a}.
They transformed the $2$-to-$1$ property of a mapping into the permutation property of another relate mapping, and thus $2$-to-$1$ mappings can be constructed from permutations.
Recently, Gao et al. \cite{gaoMto1MappingsFinite2021} generalized this to $n$-to-$1$ mapping with $\# A \equiv 0  \pmod n  \text { or } \# A \equiv 1 \pmod n$
 and found that some $n$-to-$1$ mappings can be obtained from PPs.
In \cite{yuanTwotooneMappingsInvolutions2021b}, Yuan et al. transformed the $2$-to-$1$ property of a mapping into that of another related mapping.
\begin{Lem}\label{2agw}\cite{yuanTwotooneMappingsInvolutions2021b}
Let $A, \bar{A}, S, \bar{S}$ be four finite sets and $f: A\to A,$ $g: S\to \overline{S}$, $\lambda: A\to S$ and $\overline{\lambda}: A\to\overline{S}$ be four surjective mappings  such that $\bar{\lambda} \circ f=\bar{f} \circ \lambda$.
Then $f$ is a $2$-to-$1$ mapping over $A$ if the following two conditions hold:
	\begin{enumerate}[(1)]
	\item  $\bar{f}$ is a $2$-to-$1$ mapping from $S$ to $\bar{S}$;
	\item $\sharp S$ is even, and $f$ is bijective from $\lambda^{-1}(s)=\{x \in A \mid \lambda(x)=s\}$ to $\bar{\lambda}^{-1}(\bar{f}(s))=\{x \in$ $\bar{A} \mid \bar{\lambda}(x)=\bar{f}(s)\}$ for any $s \in S .$
\end{enumerate}
\end{Lem}

Inspired by the work above, we generalize the AGW Criterion and Lemma \ref{2agw} to Theorem \ref{agwcore2}, which establishes an equivalent connection of the $n$-to-$1$ property between two mappings.
This will allow us to obtain $n$-to-$1$ mappings from other related $n$-to-$1$ mappings.
\begin{Th}
	\label{agwcore2}
Let $A, S$ and $\bar{S}$ be finite sets with $\# S=\# \bar{S}>1$ and $\# A \equiv \#S  \pmod n  $.
Let $f, g, \lambda, \bar{\lambda}$ be four mappings satisfying $\bar{\lambda} \circ f=g \circ \lambda$ (i.e. the following diagram is commutative), where $ \lambda$ and  $\bar{\lambda}$ are surjective.
\begin{equation*}
	\xymatrix{
		A \ar[rr]^{f}\ar[d]_{\lambda} &   &  A  \ar[d]^{\overline{\lambda}} \\
		S	 \ar[rr]^{g} &  & \overline{S} .}
\end{equation*}
Assume $f$ is bijective from $\lambda^{-1}(s)$ to $\bar{\lambda}^{-1}\left( g(s) \right)$,  for each $s \in S$.
Put three statements below:
	\begin{enumerate}[(1)]
	\item $f$ is an $n$-to-$1$ mapping over $A$;
	\item $g$ is an $n$-to-$1$ mapping from $S$ to $\bar{S}$;
	\item $ n \mid  \#S  $ or that $ n \nmid  \#S  $ and the exception $\bar{s_0} \in \bar{S}$ which has $t $ preimages in $S$ satisfies $\#\bar{\lambda}^{-1}(\bar{s_0})=1$, where $t=\#A \mod n$.
\end{enumerate}
Then, if (1) holds, so does (2).
If both (2) and (3) hold, so does (1).
\end{Th}
\begin{proof}
	If $g$ is $n$-to-$1$, put $b\in A$.
	We consider the number of solutions to $f(x)= b $ in $A$, where  the situation $\# f^{-1}(b)=0$ can be omitted.
	Assume $a \in A, \bar{s} \in \bar{S}, s\in S $ satisfying $f(a)=b, \bar{\lambda}(b)=\bar{s}, \lambda(a)=s$.
	One can obtain $g(s)=\bar{s}$ by $\bar{\lambda} \circ f=g \circ \lambda$.
	Since $g$ is $n$-to-$1$, except at most one $\bar{s_0}\in \bar{S}$ has $t $ preimages $s_1,s_2,...,s_{t} \in S$ of $g$, for each rest $\bar{s}\in \bar{S}$, there are exactly $n$ preimages $s_1,s_2...,s_n \in S$ of $g$.
	Clearly, $a \in \lambda^{-1}(s_i)$ for only one $i,  1\le i \le n \text{ or } t $.
	Note that $f$ is bijective from $\lambda^{-1}(s)$ to $\bar{\lambda}^{-1}\left( \bar{s_i} \right)$ for each $s\in S$.
	Thus for each $i$, there is only one element $x \in \lambda^{-1}(s_i)$ satisfying $f(x)=b$.
	In the following, the proof is divided into two situations.
	If $ n \mid  \#S  $, one can obtain that $f(x)=b$ has exactly $n$ solutions for all $b \in A$.
	If $ n \nmid  \#S  $ and the only one exception $\bar{s_0}$ which has $t $ preimages satisfies $\bar{\lambda}^{-1}(\bar{s_0})=\{b_0\}$, we obtain that $f(x)=b$ has $n$ solutions for all $b\in A$ with exactly one exception $b_0\in A$ having $t$ solutions.
	Thus, if both (2) and (3) hold, so does (1).

	If $f$ is $n$-to-$1$, 
	we set $ \bar{s} \in \bar{S} $ and pairwise different $b_1,b_2,...,b_m \in A$ satisfying $\bar{\lambda}({b_1})=\bar{\lambda}({b_2})=...=\bar{\lambda}({b_m})=\bar{s}$, where $1\le m\le \#A $.
	Then, we consider the number of solutions to $g(x)= \bar{s} $ in $S$, where the situation $\# g^{-1}( \bar{s})=0$ can be omitted.
	Since $f$ is $n$-to-$1$, 
	except at most one exception $b_{0}  \in A$ ( $i_0 \in \{1,2,...,m \}$   )  has $t $ preimages $a_{({0},1)},a_{({0},2)}...,a_{({0},t)} \in A $ of $f$, 
	for each rest ${b_i}$ ($i=1,2,...,m$), there are exactly $n$ preimages $a_{(i,1)},a_{(i,2)}...,a_{(i,n)} \in A$ of $f$ in $A$.
    If we assume $\lambda(a_{(i,1)})=\lambda(a_{(i,2)})$, then  $f$ is not bijective from $\lambda^{-1}\left(  \lambda(a_{(i,1)})    \right) $ to $\bar{\lambda}^{-1}\left( \bar{s} \right)$, since $f(a_{(i,1)} )=f(a_{(i,2)})=b_1 \in \bar{\lambda}^{-1}\left( \bar{s} \right)$.
    Thus, for each fixed $i$, $\lambda(a_{(i,j)})$ for $ 1\le j \le n \text{ or } t$ are pairwise different.
    Below, we will determine all elements in $\lambda^{-1} \left( \lambda(a_{(i,j)}) \right) $.
According to $\bar{\lambda} \circ f=g \circ \lambda$ and that $f$ is bijective from $\lambda^{-1}(s)$ to $\bar{\lambda}^{-1}\left( g(s) \right)$ for each $s \in S$, one can obtain $\#\lambda^{-1} \left( \lambda(a_{(i,j)}) \right) =   \# \bar{\lambda}^{-1}\left( \bar{s} \right) =m $.
    Note $1 \le i \le m$, and for each fixed $i$, $\lambda(a_{(i,j)})$ for all $ j $ are pairwise different.
    Thus, we have 
    $\lambda(a_{(1,j)})=\lambda(a_{(2,P_2(j))})=...=\lambda(a_{(m,P_{m}(j))})$, i.e., 
     $\lambda^{-1} \left( \lambda(a_{(1,j)}) \right)  = \{  a_{(1,j)},a_{(2,P_2(j))},...a_{(m,P_{m}(j))}       \}$, and the situation when $t>0$, $m>1$ and $b_{0}$ be one of $b_i$ can not happen, where $P_2,P_3,...P_{m}$ are permutations of $\{1,2,...,n\}$.
	This is equivalent to $ 1\le j \le t $, $$\# \{  \lambda(a_{(i,j)})   \text{ for all } i \text{ and } j \}=t$$ (resp. for  $ 1\le j \le n $,  $$\# \{  \lambda(a_{(i,j)})   \text{ for all } i \text{ and } j \}=n$$ ), 	when $t \ne 0 $, $m=1$ and $b_{0} =b_1$ (resp. all other situations).
	Thus, $g(x)=\bar{s}$ has exactly $n$ solutions ($\lambda(a_{(1,j)})$ for $ 1\le j \le n $), except for at most one exception has $t$ solutions.
	Thus, if (1) holds, so does (2).
\end{proof}

Applying Theorem \ref{agwcore2} into different forms of polynomials, e.g., $ h(\psi(x)) \phi(x)+g(\psi(x))$, $L_{1}(x)+L_{2}(x) g(L_{3}(x))$, $x^rh(x^s)$ and $ g\left(  x^{q^k} -x +\delta   \right) +cx$, we can obtain several general $n$-to-$1$ mappings.
In each form of $n$-to-$1$ constructions, we propose at least one explicit example.

\subsection{$n$-to-$1$ mappings of the form $ h(\psi(x)) \phi(x)+g(\psi(x)) $}

In this subsection, we apply Theorem \ref{agwcore2} into $n$-to-$1$ mappings of the form $ h(\psi(x)) \phi(x)+g(\psi(x)) $.

\begin{Prop}
	\label{a1}
Let $q$ be a prime power and $ m,n $ be positive integers, satisfying $n \mid q$.
Consider any polynomial $g[x] \in \mathbb{F}_{q^{m}}[x]$, any additive polynomials $\varphi, \psi \in \mathbb{F}_{q^{m}}[x]$,
any $q$-polynomial $\bar{\psi} \in \mathbb{F}_{q^{m}}[x]$ satisfying $\varphi \circ \psi=\bar{\psi} \circ \varphi$, and $ n \mid  \# \psi\left(\mathbb{F}_{q^{m}}\right)=  \# \bar{\psi}\left(\mathbb{F}_{q^{m}}\right) >1 $,
 and any polynomial $h \in \mathbb{F}_{q^{m}}[x]$ such that $h\left(\psi\left(\mathbb{F}_{q^{m}}\right)\right) \subseteq \mathbb{F}_{q}^*$.
Assume $\ker(\phi) \cap \ker(\psi)=\{0\}$ and $\dim \operatorname{ker}(\psi)=\operatorname{dim} \operatorname{ker}(\bar{\psi})$.
Then
$$f(x)=h(\psi(x)) \phi(x)+g(\psi(x))$$
is an $n$-to-$1$ mapping over $\mathbb{F}_{q^{m}}$ if and only if $\bar{f}(x)=h(x) \phi(x)+\bar{\psi} \circ g(x)$ is an $n$-to-$1$ mapping over $\psi\left(\mathbb{F}_{q^{m}}\right)$.
\end{Prop}
\begin{proof}
We can obtain $f\left(\psi^{-1}(s)\right) \subseteq \bar{\psi}^{-1}(\bar{f}(s)) $ from the following commutative diagram.
\begin{equation*}
	\xymatrix{
		\gf_{q^m} \ar[rr]^{f}\ar[d]_{\psi} &   &  \gf_{q^m}  \ar[d]^{\overline{\psi}} \\
		\psi(\gf_{q^m})	 \ar[rr]^{\bar{f}} &  & \overline{\psi}(\gf_{q^m}) .}
\end{equation*}
Then, assume $a , b \in \psi^{-1}(s)$ satisfying $f(a)=f(b)$.
After simplifying $$h(\psi(a)) \phi(a)+g(\psi(a))=h(\psi(b)) \phi(b)+g(\psi(b))$$ by $\psi(b)=s$ and since $\phi$ is additive, we get $h(s) \phi(a-b)=0$.
Note that $h\left(\psi\left(\mathbb{F}_{q^{m}}\right)\right) \subseteq \mathbb{F}_{q}^{*}$ and $\psi(a-b)=0$.
We have $ (a-b)  \in \ker (\phi) \cap \ker (\psi)$, and thus $a=b$.
Since $\dim \ker(\psi)=\dim \ker(\bar{\psi}), f$ is surjective from $\psi^{-1}(s)$ to $\bar{\psi}^{-1}(\bar{f}(s)) $.
Thus, $f$ is bijective from $\psi^{-1}(s)$ to $\bar{\psi}^{-1}(\bar{f}(s))$ for any $s \in \psi\left(\mathbb{F}_{q^{m}}\right)$.
Furthermore, $q \mid \#\psi\left(\mathbb{F}_{q^{m}}\right)$ by Lemma \ref{linear}.
According to Theorem \ref{agwcore2},  $f$ is $n$-to-$1$ if and only if $\bar{f}$ is $n$-to-$1$.
\end{proof}

It is not hard to give an example for $g(\gf_q) \subseteq \ker (\tr_{q^m/q})$ with $m \ge 2$, which will be used to propose some corollaries below.
For example,  $g=g_0^q-g_0 \ne 0$ satisfies this condition, where $g_0\in \gf_{q^m}[x]$.
\begin{Cor}
Let $a \neq 0$.
Assume $g[x] \in \gf_{q^m}[x]$ satisfying $g(\gf_q) \subseteq \ker(\tr_{q^m/q})$ and $g(0)=0$.
Then $ f(x)=a\tr_{q^m/q}(x)^d+g(\tr_{q^m/q}(x)) $ is $n$-to-$1$ if and only if $\gcd(d, q-1)=n$.
\end{Cor}
\begin{proof}
Let $\psi=\bar{\psi}=\tr_{q^m/q}$, $\phi(x)=x$, $h(x)=a x^{d}$ in Proposition \ref{a1}.
Then $\bar{f}=h$ is $n$-to-$1$ over $\gf_{q}^*$ if and only if $\gcd(d, q-1)=n$.
Note $\bar{f}(0)=0$ and $  x_0=0 $ is the only element satisfying $ \bar{\psi}^{-1}\left( f(x_0)   \right) =0$.
According to Theorem \ref{agwcore2} and the proof of Proposition \ref{a1}, $ f$ is $n$-to-$1$ if and only if $\gcd(d, q-1)=n$.
\end{proof}

%
%
%

\begin{Cor}
	Let $q=3^k$,  m be  odd.
	Assume $g\in \gf_{q^m}[x]$ satisfying $g(\gf_q) \subseteq \ker(\tr_{q^m/q})$.
	Then, $f(x)= \tr_{q^m/q}(x) x^2     -  \tr_{q^m/q}(x)  x        +    x^2-x +g(\tr_{q^m/q})$ is $3$-to-$1$ over $\gf_{q^m}$.
\end{Cor}
\begin{proof}
	Let $\psi=\bar{\psi}=\tr_{q^m/q}$, $\phi(x)=x^2-x$, $h(x)=x+1$ in Proposition \ref{a1}.
	Then $\bar{f}(x)=(x^2-x)(x+1) =x^3-x$ is $3$-to-$1$ over $\gf_{q}$, by Theorem \ref{de3p3}.
	According to Proposition \ref{a1}, $ f$ is $3$-to-$1$.
\end{proof}

\subsection{$n$-to-$1$ mappings of the form $L_{1}(x)+L_{2}(x) g\left(L_{3}(x)\right)$}

In this subsection, we apply Theorem \ref{agwcore2} into $n$-to-$1$ mappings of the form $L_{1}(x)+L_{2}(x) g\left(L_{3}(x)\right)$.

\begin{Prop}\label{hbjhvsdacax}
Let $q$ be a prime power and $ m,n $ be positive integers satisfying $n \mid q$.
Let $L_{1}(x)$, $L_{2}(x)$, $L_{3}(x) \in \mathbb{F}_{q}[x]$ be $q$-linearized, and $g(x) \in \mathbb{F}_{q^{m}}[x]$ be such that $g\left(L_{3}\left(\mathbb{F}_{q^{m}}\right)\right) \subseteq \mathbb{F}_{q} $ and $ n \mid  \# L_3(\gf_{q^m}) $,
and {$  \# L_3(\gf_{q^m})   \equiv q  \pmod n  $}.  
Assume $\ker \left(F_{y}\right) \cap \ker\left(L_{3}\right)=\{0\}$ for any $y \in L_{3}\left(\mathbb{F}_{q^{m}}\right)$, where $F_{y}(x)=L_{1}(x)+L_{2}(x) g(y)$.
Then
$$f(x)=L_{1}(x)+L_{2}(x) g\left(L_{3}(x)\right)$$
is an $n$-to-$1$ mapping over $\mathbb{F}_{q^{m}}$ if and only if $\bar{f}(x)=L_{1}(x)+L_{3}(x) g(x)$ is an $n$-to-$1$ mapping over $L_3\left(\mathbb{F}_{q^{m}}\right)$.
\end{Prop}
\begin{proof}
We can obtain $f\left(L_3^{-1}(s)\right) \subseteq L_3^{-1}(\bar{f}(s)) $ from 
\begin{equation*}
	\xymatrix{
		\gf_{q^m} \ar[rr]^{f}\ar[d]_{L_3} &   &  \gf_{q^m}  \ar[d]^{L_3} \\
		L_3(\gf_{q^m})	 \ar[rr]^{\bar{f}} &  & L_3(\gf_{q^m}). }
\end{equation*}
Then, assume $a , b \in L_3^{-1}(y)$ satisfying $f(a)=f(b)$.
After simplifying it by $  L_3(a) =y  $, we get $F_{y}(a)=F_{y}(b)$.
Note that $  F_{y}(x)     $ is additive for $x$ and $L_3(a-b)=0$.
So $(a-b) \in \ker \left(F_{y}\right) \cap \ker \left(L_{3}\right)$, that is, $a=b$.
Thus, $f$ is bijective from $L_3^{-1}(s)$ to $L_3^{-1}(\bar{f}(s))$ for any $s \in \psi\left(\mathbb{F}_{q^{m}}\right)$.
According to Theorem \ref{agwcore2},  $f$ is $n$-to-$1$ if and only if $\bar{f}$ is $n$-to-$1$.
\end{proof}

We provide a corollary with $p=3$.
%
%

\begin{Cor}
	Let $q$ be a power of $3$, and $ 3 \nmid m$.
	Then, $$f(x)=x^2+x(\tr_{q^m/q}(x)^2-\tr_{q^m/q}(x)-a)$$ is $3$-to-$1$ over $\gf_{q^m}$, where $a$ is a square in $\gf_{q}$.
\end{Cor}
\begin{proof}
	Let $L_1(x)=x^2, L_2(x)=x, L_3(x)=\tr_{q^m/q}(x)$ in Proposition \ref{hbjhvsdacax}.
	Assume $z \in \gf_q^*$ and $f_y(z)=z^2+z(y^2-y-a)=0$.
	Then we obtain $z=-y^2+y-a\in \gf_{q}$ and $\tr_{q^m/q}(z)\ne 0$, since $ 3 \nmid m$.
	Thus $\ker \left(F_{y}\right) \cap \ker\left(L_{3}\right)=\{0\}$.
	By Theorem \ref{de3p3}, $\bar{f}(x)=x^2+x(x^2-x-a)=x^3-ax$ is $3$-to-$1$ over $\gf_{q}$.
	According to Proposition \ref{hbjhvsdacax}, $f$  is $3$-to-$1$.
\end{proof}


%
%
%
%
%
%
%
%
%
%
%
%
%
%
%
%
%
%


\subsection{$n$-to-$1$ mappings of the form $x^{r} h\left(x^{s}\right)$}

In this subsection, we consider $n$-to-$1$ mappings of the form  $x^{r} h\left(x^{s}\right)$.

\begin{Prop}
	\label{mul2}
	Let $q$ be a prime power, $r, s$ be positive integers such that $s \mid (q-1), \gcd \left(r, s \right)=1$ and $ n \mid  \frac{q-1}{s}$.
	Let $f(x)=x^{r} h\left(x^{s}\right)$, where $h[x] \in \mathbb{F}_{q}[x]$ such that $h(x) \neq 0$ if $x \neq 0$, and let $\lambda(x)=x^{s}$ and $\mu_{(q-1)/s}=\left\{x \in \mathbb{F}_{q}: x^{(q-1)/s}=1\right\} $.
	Assume $f\circ\lambda=\lambda \circ g$.
	Then, $f$ is an $n$-to-$1$ mapping over $\mathbb{F}_{q}^*$ if and only if $g(x)=x^{r} h(x)^{s}$ is an $n$-to-$1$ mapping over $\mu_{(q-1)/s}$.
\end{Prop}
\begin{proof}
	We can obtain $f\left(\lambda^{-1}(s)\right) \subseteq {\lambda}^{-1}(g(s)) $ from the following commutative diagram.
	\begin{equation*}
		\xymatrix{
			\gf_{q}^* \ar[rr]^{f(x)=x^rh(x^s)}\ar[d]_{x^s} &   &  \gf_{q}^*  \ar[d]^{{x^s}} \\
			\mu_{(q-1)/s}	 \ar[rr]^{   g(x)=x^rh(x)^s           } &  & \mu_{(q-1)/s} .}
	\end{equation*}
	Since $\gcd \left(r, s \right)=1$, one can see that $f |_{\lambda^{-1}(y)}(x)=x^rh(y)$ is injective from $\lambda^{-1}(y)$ to ${\lambda}^{-1}\left( g(y) \right)$, for each $y\in \mu_{(q-1)/s}$.
Due to $\# \lambda^{-1}(y) =\# {\lambda}^{-1}\left( g(y) \right)$, $f |_{\lambda^{-1}(y)}(x)$ is surjective.
Note $n \mid \#\mu_{(q-1)/s} $.
		Then, $f$ is $n$-to-$1$ if and only if $g$ is $n$-to-$1$, according to Theorem \ref{agwcore2}.
\end{proof}
\begin{Rem} Let symbols be defined as in Proposition \ref{mul2}. 
	Clearly for any $x\in \mu_{(q-1)/s} $, $h(x) \ne 0$.
	Thus, $f^{-1}(0) = \{  0 \}$.
	And $f(x)=a$ with $ a \ne 0$ has exactly $n$ or $0$ solutions in $\mathbb{F}_{q}^*$.
	Thus, if $f$ is $n$-to-$1$ over $\mathbb{F}_{q}^*$, so does it over $\mathbb{F}_{q}$.
\end{Rem}
When $g$ is a monomial $n$-to-$1$ mapping, we have the following corollary.
\begin{Cor}
		Assume $\gcd(r,s)=1$, $ n \mid  \frac{q-1}{s}$, and $h(x) \in \gf_q [x]$ satisfying $h(y)^s=\alpha y^t$ for all $y \in \mu_{(q-1)/s}$, where $\alpha \in \mu_{(q-1)/s}$.
		Then, $x^rh(x^s)$ is $n$-to-$1$ if and only if $\gcd(r+t,(q-1)/s)=n$.
	\end{Cor}

If we choose a special $s$ in Proposition \ref{mul2} such that $\mu_{(q-1)/s}$ is exactly a set of all nonzero elements of a finite field, then the following corollary is obtained, which is to recursively obtain new $n$-to-$1$ mappings from known ones on their subfields.
\begin{Cor}
	\label{diguiuse2}
	Let $q$ be a prime power and $n,m, r$ be positive integers with $\gcd(q-1, m)=1$ and $n \mid (q-1)$.
	Let $h(x) \in \mathbb{F}_{q}[x] .$
	Then $f(x)=x^r h\left(x^{\frac{q^{m}-1}{q-1}}\right)$ is an $n$-to-$1$ mapping over $\mathbb{F}_{q^{m}}^*$ if and only if $g(x)=x^{r} h(x)^{m}$ is an $n$-to-$1$ mapping over $\mathbb{F}_{q}^*$.
\end{Cor}
\begin{proof}
	Let $ s= \frac{q^{m}-1}{q-1}$.
	Since $s=m+\sum_{i=1}^{m-1}\left(q^{i}-1\right) \equiv	m \bmod (q-1)$ and $h(x) \in \mathbb{F}_{q}[x]$,  one can obtain that $ \gcd(q-1, s)=\gcd(q-1, m)=n  $  and  $ g(x)=x^{r} h(x)^{\frac{q^{m}-1}{q-1}}=x^{r} h(x)^{m} $   for any $ x \in \gf_q ^* $.
	According to Proposition \ref{mul2}, $f$ is $n$-to-$1$ over	$\mathbb{F}_{q^{m}}$ if and only if $x^{r} h(x)^{m}$ is $n$-to-$1$ over $\mathbb{F}_{q}^*$.
\end{proof}

Below, we consider constructing $n$-to-$1$ mappings from the cyclotomic perspective.
	\begin{Prop}
	\label{piecewisegenerel}
	Let $\beta$ be a primitive element of $\mathbb{F}_{q}$ and $\omega=\beta^{s}$ be a generator of the subgroup $\mu_{\ell}$, where $s \mid (q-1)$, $n \mid \ell=\frac{q-1}{s}  $, and $\gcd \left(r, s \right)=1$.
	For any $ 0\le i \le \ell-1 $, let $a_{0}, a_{1}, \ldots, a_{2}, \ldots, a_{\ell-1}$ be $\ell$ elements in the multiset $\bigcup_{i = 1}^{n} \{*  c_1,c_2,....c_{ \ell/n } *\}  $, where $c_1,c_2,....c_{\lceil \ell/n \rceil}$ is $ \ell/n   $ distinct elements in $\{  0,1, \ldots, \ell-1 \}.$
	Assume that  $0 \leq m_{i} \leq s-1$ are integers for $ 0\le i \le \ell-1 $, and $ h(x)=\sum_{i=0}^{\ell-1} h_i x^i \in \gf_{q}[x] $ is a reduced polynomial modulo $x^{\ell}-1$, such that
	\begin{equation}
		\label{matrix}
		H=A^{-1}B,
	\end{equation}
	where
	\begin{equation*}
		\label{matrixA}
		A=\left(
		\begin{array}{cccc}
			1 & 1 & \cdots & 1 \\
			1 & \omega & \cdots & \omega^{\ell-1} \\
			1 & \omega^{2} & \cdots & \omega^{2(\ell-1)} \\
			& \cdots & \cdots & \\
			1 & \omega^{\ell-1} & \cdots & \omega^{(\ell-1)(\ell-1)}
		\end{array}
		\right)
		\text{is a Vandermonde matrix,}
	\end{equation*}
	\begin{equation*}
		H=\left(\begin{array}{c}
			h_{0} \\
			h_{1} \\
			h_{2} \\
			\vdots \\
			h_{\ell-1}
		\end{array}\right)
		\text { and }
		B=\left(\begin{array}{c}
			\beta^{\ell m_{1}+a_{1}- r}  \\
			\vdots \\
			\beta^{\ell m_{i}+a_{i}-i r} \\
			\vdots \\
			\beta^{\ell m_{(\ell-1)}+a_{(\ell-1)}- (\ell-1) r}
		\end{array}\right) .
	\end{equation*}
	Then $  f (x) = x^r h(x^s) $ is an $n$-to-$1$ mapping over $\mathbb{F}_{q}$.
\end{Prop}
\begin{proof}
	According to $ AH=B $, for $ 0\le i \le \ell-1 $, we have
	\begin{equation*}
			h\left(\omega^{i}\right) =\beta^{\ell m_{i}+a_{i}-i r}
	\end{equation*}
	Thus, for $0 \leq i \leq \ell-1$, one can obtain that
	\begin{equation*}
		\label{geq}
			g\left(\omega^{i}\right) =\omega^{i r} h\left(\omega^{i}\right)^{s}=\omega^{a_{i}},
	\end{equation*}
	i.e., $g(x)=x^{r} h(x)^{s}$ is $n$-to-$1$ over $\mu_{\ell}$.
	Clearly $n \mid \#\mu_\ell$ and $f^{-1}(0)=\{0 \}$.
	According to Proposition \ref{mul2}, $  f $ is $n$-to-$1$ over $\mathbb{F}_{q}^*$.
	Since $h(x) \ne 0$ for any $x\in \mu_\ell$,  $f $ is $n$-to-$1$ over $\mathbb{F}_{q}$.
\end{proof}

Motivated by Proposition \ref{piecewisegenerel}, we have the following explicit constructions using piecewise method.

\begin{Th}
	\label{miu2}
	Let $q$ be an odd prime power satisfying $q \equiv 3 \pmod 4 $, and
	$$ f(x)=\frac{a-b}{2} x^{\frac{q-1}{2}+r}+\frac{a+b}{2} x^{r} \in \gf_q[x],$$
	where $ a,b \in \gf_q  $.
	Assume that $ s=\frac{q-1}{2}  $ and $\gcd \left(r, s \right)=1$.
	Then $f$ is a $2$-to-$1$ mapping over $\mathbb{F}_{q}$ if and only if
	\begin{enumerate}[(1)]
		\item $a \in S_1 $,  $b\in S_{(-1)^r}  $; or
		\item $a \in S_{-1} $,  $b\in S_{(-1)^{r+1}}  $,
	\end{enumerate}
where, $S_1$ (resp. $S_{-1}$) are the set containing all squares (resp. non-squares) of $\mathbb{F}_{q}^*$.
\end{Th}
\begin{proof}
	As we know,	$h(x)=\frac{a-b}{2} x^{\frac{q-1}{2}}+\frac{a+b}{2}$,  and $ h(1)=a, h(-1)=b  $.
	Then, $f(x)$ can be rewritten as
	\begin{equation}
		\label{twoelements}
		f(x)=x^{r} h\left(x^{s}\right)=\left\{
		\begin{array}{ll}
			0, & x=0 \\
			a x^{r}, & x \in S_{1} \\
			b x^{r}, & x \in S_{-1}. \\
		\end{array}\right.
	\end{equation}
Put $g(x)=x^rh(x)^s$ over $\mu_2$.
Note $ S_{ 1}=\left\{y \in \mathbb{F}_{q}^* |  \  y^{\frac{q-1}{2}}=  1 \right\}$ and $ S_{-1}=\left\{y \in \mathbb{F}_{q}^* |  \  y^{\frac{q-1}{2}}= - 1 \right\}$.
If $a \in S_1 $ and $b\in S_{(-1)^r}  $, we have $g(1)=a^{(q-1)/2}=1$ and $g(-1)=(-1)^r b^{(q-1)/2}=1$, i.e., $g(x)$ is $2$-to-$1$ over $\mu_2$.
If $a \in S_{-1} $ and  $b\in S_{(-1)^{r+1}}  $, we have $g(1)=a^{(q-1)/2}=-1$ and $g(-1)=(-1)^r b^{(q-1)/2}=-1$, i.e., $g(x)$ is $2$-to-$1$ over $\mu_2$.
Clearly $2 \mid \#\mu_2$.
According to Proposition \ref{mul2},  $f$ is $2$-to-$1$ over $\gf_q^*$.
By $f^{-1}(0)=\{0 \}$, $f$ is $2$-to-$1$ over $\gf_q$.

Conversely, if $f$ is $2$-to-$1$, $g(x)$ being $2$-to-$1$ over $\mu_2$ can only have two situations: $g(1)=g(-1)=1$ or $g(1)=g(-1)=-1$.
From the deduction above, the result is established.
\end{proof}

We represent an example of Corollary \ref{diguiuse2} by using Theorem \ref{miu2}.
Obtaining examples by Theorems \ref{miu3} and \ref{nmiu3} are similar with Example \ref{diguilizi}, and thus we omitted them.
\begin{Exa}\label{diguilizi}
	Let $t,k$ be positive integers, $q=p^k$ and  $ g(x)=x^{r}  \left(   \frac{a-b}{2} x^{\frac{q-1}{2}}+\frac{a+b}{2}     \right) $ be a $2$-to-$1$ mapping over $\mathbb{F}_{q}$ as  in Theorem \ref{miu2}.
	Then, $$f(x)=x^r       \left(   \frac{a-b}{2} x^{\frac{q^{p^t}-1}{2}}+\frac{a+b}{2}     \right)^{1/p^t}           $$ is $2$-to-$1$ over $\mathbb{F}_{q^{p^t}}^*$.
\end{Exa}
\begin{proof}
	Let $m=p^t$ in Corollary \ref{diguiuse2}.
	Then $h(x)=\left(   \frac{a-b}{2} x^{\frac{q-1}{2}}+\frac{a+b}{2}     \right)^{1/p^t} $.
	By Corollary \ref{diguiuse2}, $f(x)=x^r h\left(x^{\frac{q^{   p^t    }-1}{q-1}}\right) $ is $2$-to-$1$ over $\mathbb{F}_{q^{p^t}}^*$.
\end{proof}

\begin{Th}
	\label{miu3}
	Let $q$ be a prime power such that $s=\frac{q-1}{3}$ is an integer, $ \alpha $ be a primitive element of $ \gf_q $, and $\gcd \left(r, s \right)=1$.
	Assume that $ g(x)=x^rh(x)^s $ is over $ \mu_{3} $.
	Let $$ h(x)=\frac{b-c-a \omega^{2}+b \omega^{2}}{(-1+\omega)^{2} \omega} x^{2}+     \frac{c+a \omega-b(1+\omega)}{(-1+\omega)^{2} \omega(1+\omega)}x+     \frac{c+a \omega^{3}-b \omega(1+\omega)}{(-1+\omega)^{2}(1+\omega)}    ,$$
	where $ a,b,c \in \gf_q  $, $\omega=\alpha^{\frac{q-1}{3}}$.
	Then $f(x)=x^rh(x^s)$ is a $3$-to-$1$ mapping over $\mathbb{F}_{q}$ if and only if
	\begin{enumerate}[(1)]
		\item $a \in S_1 , b\in S_{\omega^{-r}}, c \in S_{\omega^{-2r}} $; or
		\item $a \in S_{\omega} , b\in S_{\omega^{1-r}}, c \in S_{\omega^{1-2r}} $; or
		\item $a \in S_{\omega^2} , b\in S_{\omega^{2-r}}, c \in S_{\omega^{2-2r}} $,
	\end{enumerate}
where $ S_{i}=\left\{y \in \mathbb{F}_{q}^* | y^{\frac{q-1}{3}}=i \right\} $, $i \in  \{   1,\omega, \omega^2  \}$.
\end{Th}
\begin{proof}
	It is easy to obtain that	$ h(1)=a, h(\omega)=b, h(\omega^2)=c  $.
	Then, $f(x)$ can be rewritten as
	\begin{equation}
	\label{threeelements}
	f(x)=x^{r} h\left(x^{s}\right)=\left\{
	\begin{array}{ll}
		0, & x=0 \\
		h(1) x^{r}, & x \in S_{1} \\
		h(\omega) x^{r}, & x \in S_{\omega} \\
		h(\omega^2) x^{r}, & x \in S_{\omega^2}.
	\end{array}\right.
\end{equation}
%
%
%
	If $a \in S_1 , b\in S_{\omega^{-r}}, c \in S_{\omega^{-2r}} $, we have $g(1)=a^{(q-1)/2}=1$, $g(-1)=(\omega)^r b^{(q-1)/2}=1$ and $g(w^2)=\omega^{2r}c^\frac{q-1}{3}=1$.
	If $a \in S_{\omega} , b\in S_{\omega^{1-r}}, c \in S_{\omega^{1-2r}} $, we have $g(1)=a^{(q-1)/2}=\omega$, $g(-1)=(\omega)^r b^{(q-1)/2}=\omega$ and $g(w^2)=\omega^{2r}c^\frac{q-1}{3}=\omega$.
	If $a \in S_{\omega^2} , b\in S_{\omega^{2-r}}, c \in S_{\omega^{2-2r}} $, we have $g(1)=a^{(q-1)/2}=\omega^2$, $g(-1)=(\omega)^r b^{(q-1)/2}=\omega^2$ and $g(w^2)=\omega^{2r}c^\frac{q-1}{3}=\omega^2$.
	For these situations, $g(x)$ is always $3$-to-$1$ over $\mu_3$.
	Clearly $3 \mid \#\mu_3$.
	According to Proposition \ref{mul2} and $f^{-1}(0)=\{0 \}$,  $f$ is $3$-to-$1$.
	
	Conversely, if $f$ is $3$-to-$1$, $g(x)$ being $3$-to-$1$ over $\mu_3$ can only have these three situations above, since $\# \mu_3 =3$.
	Thus, the result is established.
\end{proof}

\begin{Th}
	\label{nmiu3}
	Let $q$ be an odd prime power such that $s=\frac{q-1}{4}$ is an integer, $ \alpha $ be a primitive element of $ \gf_q $, and $\gcd \left(r, s \right)=1$.
	Assume that $ g(x)=x^rh(x)^s $ is over $ \mu_{4} $.
	Let
\begin{equation*}
		\begin{aligned}
			h(x) =& \frac{ \left(-a \omega^3+b \left(\omega^2+\omega+1\right) \omega-c \left(\omega^2+\omega+1\right)+d\right)}{2 \omega - 2 \omega^3}x^3   \\
			&	+\frac{ \left(a -b \left(1+\omega^3+\omega\right)+c \left(\omega^3+\omega+1\right)-d\right)}{2 - 2 \omega^2}x^2     \\
			&\frac{ \left(-a \omega+b \left(\omega^3+\omega+1\right) \omega^2-c \left(\omega^3+\omega^2+1\right)+d\right)}{-2 \omega + 2 \omega^3}x \\
			&\frac{a \omega^2-b \left(\omega^2+\omega+1\right) \omega^3+c \left(\omega^2+\omega+1\right) \omega-d}{-2 + 2 \omega^2} ,
		\end{aligned}
	\end{equation*}
	where $ a,b,c, d \in \gf_q  $, $\omega=\alpha^{\frac{q-1}{4}}$.
	Then $f(x)=x^rh(x^s)$ is a $4$-to-$1$ mapping over $\mathbb{F}_{q}$ if and only if
\begin{enumerate}[(1)]
	\item $a \in S_1 , b\in S_{\omega^{-r}}, c \in S_{\omega^{-2r}},d \in S_{\omega^{-3r}} $; or
	\item $a \in S_{\omega} , b\in S_{\omega^{1-r}}, c \in S_{\omega^{1-2r}}, d \in S_{\omega^{1-3r}} $; or
	\item $a \in S_{\omega^2} , b\in S_{\omega^{2-r}}, c \in S_{\omega^{2-2r}} , d \in S_{\omega^{2-3r}}$,
	\item $a \in S_{\omega^3} , b\in S_{\omega^{3-r}}, c \in S_{\omega^{3-2r}} , d \in S_{\omega^{3-3r}}$,
\end{enumerate}
where $ S_{i}=\left\{y \in \mathbb{F}_{q}^* | y^{\frac{q-1}{4}}=i \right\} $, $i \in  \{   1,\omega, \omega^2, \omega^3  \}$.
\end{Th}
\begin{proof}
	It is not hard to derive that	$ h(1)=a, h(\omega)=b, h(\omega^2)=c , h(\omega^3)=d $.
	and $f(x)$ can be rewritten as
	\begin{equation}
		\label{}
		f(x)=x^{r} h\left(x^{s}\right)=\left\{
		\begin{array}{ll}
			0, & x=0 \\
			a x^{r}, & x \in S_{1} \\
			b x^{r}, & x \in S_{\omega} \\
			c x^{r}, & x \in S_{\omega^2}, \\
			d x^{r}, & x \in S_{\omega^3},
		\end{array}\right.
	\end{equation}
The rest proof is omitted, since it is similar with that in Theorem \ref{nmiu3}.
\end{proof}


\subsection{$n$-to-$1$ mappings of the form $ g\left(  x^{q^k} -x +\delta   \right) +cx$}

In \cite{zheng2019two}, Zheng et al. investigated the permutation property between polynomials $ f(x) = g\left(x^{q^k} - x + \delta\right) +c x $ and $ h(x) = g(x)^{q^k} - g(x) + cx +(1-c)\delta $. 
In \cite{yuanTwotooneMappingsInvolutions2021b}, Yuan et al. investigated the $2$-to-$1$ property between them.
In this subsection, we generalize above works to $n$-to-$1$ property, as an application of Theorem \ref{agwcore2}.
We then propose three explicit constructions of $3$-to-$1$ mappings over fields with character $3$ and two explicit constructions of $2^l$-to-$1$ mappings over binary fields.

\begin{Th}
	\label{Zcriterion}
	Let $ \gf_{q^m} $ be the degree $ m $ extension of the finite field $ \gf_q $ and $ \delta \in \gf_{q^m} $, $g(x)\in\gf_{q^m}[x]$, $n \mid q^m$.
	Then, $ f(x) = g\left(x^{q^k} - x + \delta\right) +c x $ is an $n$-to-$1$ mapping over $ \gf_{q^m} $ if and only if $ h(x) = g(x)^{q^k} - g(x) + cx +(1-c)\delta $ is an $n$-to-$1$ mapping over $S_{\delta} =\left\{x^{q^k}-x+\delta \  | \  x \in \gf_{q^m}\right\} $, where $ c \in \gf_{q^\ell}^* $, $ k $ is an integer with $ 1 \le k \le m-1$, $\ell=\gcd(k,m)$.
	Furthermore, if $c=1$ and $f$ is $n$-to-$1$ for any $ \delta \in \gf_{q^m} $, then $h$ is also $n$-to-$1$ over $ \gf_{q^m} $.
\end{Th}
\begin{proof}
	It is clear that $ \phi(x)=x^{q^i}-x+\delta  $ is surjective on $S_{\delta} \subset \gf_{q^m}$.
	We can obtain $f\left(\phi^{-1}(s)\right) \subseteq \bar{\phi}^{-1}({f}(s)) $ from 
\begin{equation*}
	\xymatrix{
		\gf_{q^m} \ar[rr]^{f}\ar[d]_{\phi} &   &  \gf_{q^m}  \ar[d]^{\phi} \\
		S_{\delta}	 \ar[rr]^{h} &  & S_{\delta}. }
\end{equation*}
In the above diagram, $\phi(x)=x^{q^k}-x+\delta$.
Due to $x^{q^k}-x$ is additive, $q \mid \#S_{\delta}$.
	Then, assume $a , b \in \phi^{-1}(s)$ satisfying $f(a)=f(b)$.
	We can obtain $a=b$ directly.
	Clearly, $ f$ is surjective from $\phi^{-1}(s)$ to $\bar{\phi}^{-1}(\bar{f}(s)) $.
	Thus, $f$ is bijective from $\phi^{-1}(s)$ to $\bar{\phi}^{-1}(h(s))$ for each $s \in \phi\left(\mathbb{F}_{q^{m}}\right)$.
	According to Theorem \ref{agwcore2},  $f$ is $n$-to-$1$ if and only if $h$ is $n$-to-$1$.
	
	Furthermore,  we have $ \bigcup_{\delta \in \gf_{q^m}} S_{\delta} =\gf_{q^m} $.
	Thus, if $c=1$ and $f$ is $n$-to-$1$ for any $ \delta \in \gf_{q^m} $, then $h$ so does over $ \gf_{q^m} $.
\end{proof}

Below, we use Theorem \ref{Zcriterion} to construct three explicit $3$-to-$1$ mappings. 
\begin{Prop}\label{gouzao1} 
		Let $q_1$ be a power of $3$, and $q=q_1^{m} $ satisfying $\gcd\left(\frac{3q_1 q-1}{2}, q-1\right)=1$, where $m$ is a positive integer.				
Then, $$f(x) = \left(      x^{q} -x+ \delta  \right)^{6q_1}   +x      $$
	is a $3$-to-$1$ mapping over $ \gf_{q^2} $ if and only if
	$\tr_{q^2/q}(\delta)$ is not a square in $\gf_{q}$.
\end{Prop}

\begin{proof}
	According to Theorem \ref{Zcriterion}, $f$ is $3$-to-$1$
	if and only if 	$ h(x) =x^{6q_1q} - x^{6q_1} + x  $ is a $3$-to-$1$ mapping over $S_{\delta} =\left\{z^{q}-z+\delta \  | \  z \in \gf_{q^2}\right\}=\{ z  \in \gf_{q^2}  \   |  \  \tr_{q^2 / q} (z) =\tr_{q^2 / q}(\delta)    \} $.
	It suffices to prove for any $x \in S_\delta$,	
	\begin{equation} \label{nsjciahwbsh}
	h(x+y)=h(x)
	\end{equation}
	has exactly three distinct solutions for $y$ in $S_{0}$.
	After plugging $ h(x) =x^{6q_1q} - x^{6q_1} + x  $ into, expanding and simplifying Eq. (\ref{nsjciahwbsh}), one can obtain
	\begin{equation}\label{nbsjcuwgwnd}
			y^{6q_1q}-(xy)^{ 3q_1q } - y^{6q_1}+(xy)^{3q_1} +y =0 .
	\end{equation}
Let $X=x^{q_1}$ and $ Y=y^{q_1}$.
Then Eq. (\ref{nbsjcuwgwnd}) becomes
	\begin{equation}\label{aksdbakhsbva}
		Y^{6q}-X^{3q}Y^{3q} - Y^{6}+X^3Y^3 +y =0.
	\end{equation}
	Raising both sides of Eq. (\ref{aksdbakhsbva}) to the power of $q$, one can obtain
	\begin{equation}\label{sadfsdxfrgga}
		Y^{6 }-X^3Y^3 - Y^{6q}+X^{3q}Y^{3q} +y^{q} =0.
	\end{equation}
	By summing Eq. (\ref{aksdbakhsbva})  and Eq. (\ref{sadfsdxfrgga}), 
	 $y    +y^{q} =0$
is verified.
Then, $Y    +Y^{q} =0$ is obtained.
For reducing the degree, plugging $Y^{q} =-Y$ and $Y=y^{q_1}$ into Eq. (\ref{aksdbakhsbva}), one can obtain
	$$      X^{3q} y^{3q_1} +X^3y^{3q_1} +y  =0 .   $$	
Clearly $y=0$ is a solution of Eq. (\ref{nsjciahwbsh}).
Then we assume $y \ne 0$ below.
It is easy to obtain
	$$    (  X^{3q} +X^3   )         y^{3q_1}   = -y.$$	
Note $X\in S_{\delta}$.
We have $\tr_{q^2 / q}(\delta) \ne 0$ and
	\begin{equation}\label{bhbajhsbkjhbxasd}
		\frac{-y^{3q_1}}{y}=\frac{1}{\left(\tr_{q^2 / q}(\delta)\right)^{3q_1}}.
	\end{equation}
In order to find all suitable conditions for $\delta$, we plug $y=-y^{q}$ into the numerator of the left side of Eq. (\ref{bhbajhsbkjhbxasd}).
Than we obtain
\begin{equation}\label{bjhbibiusn}
\frac{y^{3q_1q} }{y}=	y^{3q_1 q-1}=\frac{1}{\left( \tr_{q^2/q}(\delta)\right)^{3q_1}}.
\end{equation}
	Since $\gcd \left(\frac{3q_1 q-1}{2}, q-1\right)=1$, assume integers $a,b$ satisfying $\frac{3q_1 q-1}{2} a+(q-1) b=1$.
	Raising both sides of Eq. (\ref{bjhbibiusn}) to the power of $a$, we have
		\begin{equation}\label{bkhvbiusn}
			 y^{2}  =\frac{1 }{ ( \tr_{q^2/ q}(\delta) )^{3q_1a}}.
		\end{equation}	
	Note that $a$ is odd and $y^{2} \in \gf_q$.
	If $\tr_{q^2/q}(\delta)$ is a square in $\gf_{q}$, one can obtain that all solutions to Eq. (\ref{bkhvbiusn}) are in $\gf_q$.
	This will lead to $y=0$  by $y    +y^{q} =0$, a conflict.
	If $\tr_{q^2/q}(\delta)$ is not a square in $\gf_{q}$, we have
	$\tr_{q^2 / q}(\delta)=\alpha^{t\left(q+1\right)}$
	exactly for $t$ being odd, where $\alpha$ is a primitive element in $\mathbb{F}_{q^2}$.
	Then $y^{2}=\alpha^{-3q_1 a t\left(q+1\right)}$ and $y=\pm \alpha^{(-3q_1 a t) \frac{q+1}{2}}$.
	Finally, we verify that all solutions are in $S_0$.
	We have $\left(\alpha^{(-3q_1 a t) \frac{q+1}{2}}\right)^{q}=\alpha^{(-3q_1 a t)^{\frac{q^2-1+q+1}{2}}}=(-1)^{-3q_1 a t} \alpha^{(-3q_1 a t)^{\frac{q+1}{2}}}$.
	Thus $\tr_{q^2 / q}(y)=0$.
	
	In summary, $f(x)$ is $3$-$1$ if and only if $\tr_{q^2 / q}(\delta)$ is not a square in $\mathbb{F}_{q}$.
\end{proof}
\begin{Rem}
When $q=q_1, \gcd\left(\frac{3q_1 q-1}{2}, q-1\right)=1$ always holds.
Then, $f(x) = \left(      x^{q_1} -x+ \delta  \right)^{6q_1}   +x      $
is a $3$-to-$1$ mapping over $ \gf_{ q_1^2  } $ if and only if $\tr_{q_1^2 /q_1}(\delta)$ is not a square in $\gf_{q_1 }$.
\end{Rem}
\begin{Exa}
	Let $q=q_1=3^3$. 
	According to Proposition \ref{gouzao1},  $f(x) = \left(      x^{27} -x+ \delta  \right)^{162}   +x      $
	is a $3$-to-$1$ mapping over $ \gf_{3^6} $ if and only if $\tr_{3^6/3^3}(\delta)$ is not a square in $\gf_{3^3}$.
    By a Magma program searching, 
	$ \{  \beta^{2i+1}  ,   \text{ for }  i=0,1,...,12 \} $
	is the set containing all non-squares in $\gf_{3^3}$, and $f$ is $3$-to-$1$ exactly when $ \tr_{3^6 / 3^3}(\delta) $ in this set, where $\beta$ is a primitive element of $\gf_{3^3}$.
\end{Exa}

\begin{Prop}\label{gouzao2}
	Let $q_1$ be a power of $3$, and $q=q_1^{m} $ satisfying $\gcd \left(\frac{3q_1 q-1}{2}, q-1\right)=1$, where $m$ is a positive integer.		
Then, $$f(x) = \left(      x^{q} -x+ \delta  \right)^{3q_1+1}   +x      $$
is a $3$-to-$1$ mapping over $ \gf_{q^2} $ if and only if
$\tr_{q^2 / q}(\delta) \ne 0$ and $    \frac{ \tr_{q^2/q}(\delta)^{3q_1}   -1}{ \tr_{q^2/q}(\delta) }   $ is not a square in $\gf_{q}$.
\end{Prop}
\begin{proof}
According to Theorem \ref{Zcriterion}, $f$ is $3$-to-$1$
if and only if 	$ h(x) = x^{3q_1  q+q} -x^{3q_1+1}+ x $ is a $3$-to-$1$ mapping over $S_{\delta} =\left\{z^{q}-z+\delta \  | \  z \in \gf_{q^2}\right\}=\{ z  \in \gf_{q^2}  \   |  \  \tr_{q^2 / q} (z) =\tr_{q^2 / q}(\delta)    \} $.
It suffices to prove for any $x \in S_\delta$,	
\begin{equation} \label{acsaqwefqsdvac}
	h(x+y)=h(x)
\end{equation}
has exactly three distinct solutions for $y$ in $S_{0}$.
Plug $ h(x) = x^{3q_1  q+q} -x^{3q_1+1}+ x $ into, and then expanding and simplifying Eq. (\ref{acsaqwefqsdvac}), one can obtain
\begin{equation}\label{bhsjhchwgwgbsh}
	x^{q} y^{3q_1 q}+x^{3q_1 q} y^{q}+y^{3q_1 q+q}-y x^{3q_1}-x y^{3q_1}-y^{3q_1+1}+y=0.
\end{equation}
Let $X=x^{3q_1}$ and $Y=y^{3q_1}$.
Then Eq. (\ref{bhsjhchwgwgbsh}) is reduced to
\begin{equation}\label{asdasbcaisv}
x^{q} Y^{ q}+X^{ q} y^{q}+y^{q}     Y^{ q}          -y X-x Y-yY+y=0.
\end{equation}
	Raising both sides of Eq. (\ref{asdasbcaisv}) to the power of $q$, we have
\begin{equation}\label{vadsgqregv}
    x Y+X y+y Y         -y^{ q} X^{ q}-x^{ q} Y^{ q}-y^{ q}Y^{ q}+y^{ q}=0
\end{equation}
By summing Eq. (\ref{asdasbcaisv})  and Eq. (\ref{vadsgqregv}), we can verify $y    +y^{q} =0$ and get $Y    +Y^{q} =0$.
Plugging $Y^{q} =-Y$ and $Y=y^{3q_1}$ into Eq. (\ref{asdasbcaisv}), one can obtain
\begin{equation}\label{asdvsgqefaxvxcb}
  - x^{q} y^{3q_1}-X^{ q}y            -y X-x y^{3q_1}+y=0 .
\end{equation}
Note $X \in S_{0}$, after simplifying Eq. (\ref{asdvsgqefaxvxcb}), we get
\begin{equation}\label{acsvsdvadfsd}
- \tr_{q^2/q}(\delta) y^{3q_1} =    \tr_{q^2/q}(\delta)^{3q_1}y   -y.
\end{equation}
Plug $y=-y^{q}$ into the left side of Eq. (\ref{acsvsdvadfsd}), we obtain
\begin{equation}\label{oiuhijhbjhhbbbbbs}
	 \tr_{q^2/q}(\delta) y^{3q_1q} =    \tr_{q^2/q}(\delta)^{3q_1}y   -y.
\end{equation}
Clearly $y=0$ is a solution of Eq. (\ref{acsaqwefqsdvac}).
We assume $y \ne 0$ below.
Clear $\tr_{q^2 / q}(\delta) \ne 0$ is necessary for $f$ being $3$-to-$1$.     
Otherwise Eq. (\ref{oiuhijhbjhhbbbbbs}) has no nonzero solution.
Then, we have
\begin{equation}\label{nbsjnsjhshsb}
y^{3q_1q-1} =   \frac{ \tr_{q^2/q}(\delta)^{3q_1}   -1}{ \tr_{q^2/q}(\delta) } .
\end{equation}
Assume integers $a,b$ satisfying $\frac{3q_1 q-1}{2} a+(q-1) b=1$. Note $a$ is odd.
Raising both sides of Eq. (\ref{nbsjnsjhshsb}) to the power of $a$, we have
\begin{equation}\label{nkajsbiviyvh}
	y^{2}  =  \left(  \frac{ \tr_{q^2/q}(\delta)^{3q_1}   -1}{ \tr_{q^2/q}(\delta) }   \right) ^a      .
\end{equation}	
If $ \frac{ \tr_{q^2/q}(\delta)^{3q_1}   -1}{ \tr_{q^2/q}(\delta) } $ is a square in $\gf_{q}$, one can obtain all solutions to Eq. (\ref{bkhvbiusn}) are in $\gf_q$.
This will lead to $y=0$, a conflict.
If $ \frac{ \tr_{q^2/q}(\delta)^{3q_1}   -1}{ \tr_{q^2/q}(\delta) } $ is not a square in $\gf_{q}$, we have
$ \frac{ \tr_{q^2/q}(\delta)^{3q_1}   -1}{ \tr_{q^2/q}(\delta) } =\alpha^{t\left(q+1\right)}$
, where $\alpha$ is a primitive element in $\mathbb{F}_{q^2}$ and $t$ is odd.
Then $y^{2}=\alpha^{\left(q+1\right)a t}$ and $y=\pm \alpha^{ \frac{q+1}{2}a t}$.
One can easily verify $ \pm \alpha^{ \frac{q+1}{2}a t} \in S_0$.

In summary, $f(x)$ is $3$-to-$1$ if and only if   $\tr_{q^2 / q}(\delta) \ne 0$ and $    \frac{ \tr_{q^2/q}(\delta)^{3q_1}   -1}{ \tr_{q^2/q}(\delta) }   $ is not a square in $\gf_{q}$.
\end{proof}
\begin{Exa}
	Let $q=q_1=3^3$. 
	According to Proposition \ref{gouzao2},  $f(x) = \left(      x^{27} -x+ \delta  \right)^{82}   +x      $
	is a $3$-to-$1$ mapping over $ \gf_{3^6} $ if and only if 
	$\tr_{3^6/3^3}(\delta) \ne 0$ and $    \frac{ \tr_{3^6/3^3}(\delta)^{81}   -1}{ \tr_{3^6/3^3}(\delta) }   $ is not a square in $\gf_{3^3}$.
	By a Magma program searching, 
	$\{
	{\beta^4,\beta^5,  \beta^7, \beta^8, \beta^{10}, \beta^{11}, \beta^{12}, 2, \beta^{15} ,\beta^{19}, \beta^{20},  \beta^{21},  \beta^{24}  }
	 \}$
	is the set for $ \tr_{3^6 / 3^3}(\delta) $ exactly satisfying that $    \frac{ \tr_{3^6/3^3}(\delta)^{81}   -1}{ \tr_{3^6/3^3}(\delta) }   $ is  a non-square in $\gf_{3^3}$, where $\beta$ is a primitive element of $\gf_{3^3}$.
	 And a Magma program verified that $f$ is $3$-to-$1$ exactly when $ \tr_{3^6 / 3^3}(\delta) $ is in this set.
\end{Exa}

\begin{Prop}\label{gouzao3}
	Let $q$ be a power of  $ 3 $.
Then,
	$$f(x) = \left(      x^{q} -x+ \delta  \right)^{2q+1}   +x      $$
is a $3$-to-$1$ mapping over $ \gf_{q^2} $ if and only if
$ (     \tr_{q^2 / q}(\delta)       )^2    -1   $ is not a square in $\gf_{q}$.
\end{Prop}
\begin{proof}
	According to Theorem \ref{Zcriterion}, $f$ is $3$-to-$1$
if and only if 	$ h(x) = x^{2q^2+q} -x^{2 q+1}+ x $ is a $3$-to-$1$ mapping over $S_{\delta} =\left\{z^{q}-z+\delta \  | \  z \in \gf_{q^2}\right\}=\{ z  \in \gf_{q^2}  \   |  \  \tr_{q^2 / q} (z) =\tr_{q^2 / q}(\delta)    \} $.
It suffices to prove for any $x \in S_\delta$,	
\begin{equation} \label{avdefwefqwdasd}
	h(x+y)=h(x)
\end{equation}
has exactly three distinct solutions for $y$ in $S_{0}$.
Plug $ h(x) = x^{2q^2+q} -x^{2 q+1}+ x $ into, and then expanding and simplifying Eq. (\ref{avdefwefqwdasd}), one can obtain
\begin{equation}\label{bshxcjhxjsn}
		\begin{aligned}
			&x^q y^{2q^2}-x^{q^2} y^{q^2+q}+y^q x^{2q^2}-x^{q^2+q} y^{q^2}+y^{2q^2+q} \\
			+&x^{q_1} y^{q_1+1}+x^{q_1+1} y^{q_1}-y x^{2 q_1}-x y^{2  q_1}-y^{2  q_1+1}+y =0. \\
		\end{aligned}
\end{equation}
Let $X=x^{q}$ and $ Y=y^{q}$.
Then Eq. (\ref{bshxcjhxjsn}) becomes
\begin{equation}\label{vajsjxchwg}
	\begin{aligned}
		&x^q Y^{2 q}-X^{q} y^q Y^{q}+y^q X^{2q}- x^q  X^{ q} Y^{ q} +   y^q  Y^{2 q}+X y  Y     +x X Y-y X^{2}-x Y^{2}-yY^{2}+y =0 . \\
	\end{aligned}
\end{equation}
Raising both sides of Eq. (\ref{vajsjxchwg}) to the power of $q$, we have
\begin{equation}\label{aoihsnciouba9sbxu}
	\begin{aligned}
	&x Y^{2 }-X y Y+y X^{2}- x  XY +   y  Y^{2}               +X^q y^q  Y^q  +x^q X^q Y^q-y^q X^{2q}-x^q Y^{2q}-y^qY^{2q}+y^q =0. \\
\end{aligned}
\end{equation}
By summing Eq. (\ref{vajsjxchwg})  and Eq. (\ref{vajsjxchwg}), $y    +y^{q} =0$ is verified.
Plugging $y^{q} =-y$ and $Y^{q} =-Y$ into Eq. (\ref{vajsjxchwg}), we have
\begin{equation}\label{sdfefxzxcasdFEfgv}
		 yY^{2}         +   (  x^q -x)   Y^{2 }  +  (X -X^{q}   ) y  Y             +  (x X     + x^q  X^{ q} )Y  =   y X^{2q} +y X^{2}  -y.
\end{equation}
Substituting $Y=-y$ into Eq. (\ref{vajsjxchwg}), we obtain
\begin{equation}\label{}
	y^{3}                =    x^{2}y + X^{2}y -xXy - y.
\end{equation}
Note $x,X \in S_{\delta}$. One can get
\begin{equation}\label{}
	y^{3}                =    \tr_{q^2/q}(\delta) ^2y- y.
\end{equation}
Clearly $y=0$ is a solution.
If $  \tr_{q^2/q}(\delta) - 1 $ is a square in $\gf_{q}$, we obtain $y^q=y$ and $y=0$ is the only one solution to Eq. (\ref{avdefwefqwdasd}).
If $ \tr_{q^2/q}(\delta) - 1 $ is not a square in $\gf_{q}$, we have
$ \tr_{q^2/q}(\delta) - 1=\alpha^{t\left(q+1\right)}$
, where $\alpha$ is a primitive element in $\mathbb{F}_{q^2}$ and $t$ is odd.
Then $y=\pm \alpha^{\frac{q+1}{2} t }$.
One can easily verify $\pm \alpha^{\frac{q+1}{2} t } \in S_0$.
Thus $y=\pm \alpha^{\frac{q+1}{2} t }, 0$ are all three distinct solutions to Eq. (\ref{avdefwefqwdasd}).

In summary, $f(x)$ is $3$-to-$1$ if and only if $\tr_{q^2/q}(\delta)^2 - 1$ is not a square in $\mathbb{F}_{q}$.
\end{proof}
\begin{Exa}
	Let $q=q_1=3^3$. 
	According to Proposition \ref{gouzao3},  $f(x) = \left(      x^{27} -x+ \delta  \right)^{55}   +x      $
	is a $3$-to-$1$ mapping over $ \gf_{3^6} $ if and only if 
	$ \tr_{3^6 / 3^3}(\delta) ^2 - 1$ is not a square in $\gf_{3^3}$.
	By a Magma program searching, 
	$\{
{ \beta^2,  \beta^{15}, \beta^{17}, \beta^{18}, \beta^{23},  \beta^{25} }
	\}$
	is the set for $ \tr_{3^6 / 3^3}(\delta) $ exactly satisfying that $ \tr_{3^6 / 3^3}(\delta) ^2 - 1$ is  not a square in $\gf_{3^3}$, where $\beta$ is a primitive element of $\gf_{3^3}$.
	And a Magma program verified that $f$ is $3$-to-$1$ exactly when $ \tr_{3^6 / 3^3}(\delta) $ is in this set.
\end{Exa}

\begin{Prop}\label{2gouzao1}
	Let $q_1$ be a power of  $ 2 $, $q=q_1^m$, where $m$ is a positive integer.  
Then,
$$f(x) =                      \left(      x^{q} +x+ \delta  \right)^{q_1+1}   +x      $$
is a $q_1$-to-$1$ mapping over $ \gf_{q^2} $ if and only if 
$\tr_{q^2/q}(\delta)\ne 0$ and $1+\frac{1}{\tr_{q^2/q}(\delta)}  $ is the $(q_1-1)$-power of an element in $\gf_q^*$.
\end{Prop}
\begin{proof}
According to Theorem \ref{Zcriterion}, $f$ is $q_1$-to-$1$
if and only if 	$ h(x) =     x^{q(q_1+1)} +x^{q_1+1}    +x    $ is a $q_1$-to-$1$ mapping over $S_{\delta} =\left\{z^{q}+z+\delta \  | \  z \in \gf_{q^2}\right\}=\{ z  \in \gf_{q^2}  \   |  \  \tr_{q^2 / q} (z) =\tr_{q^2 / q}(\delta)    \} $.
It suffices to prove for any $x \in S_\delta$,	
\begin{equation} \label{asdaavqefqwdasdasd}
	h(x+y)=h(x)
\end{equation}
has exactly $q_1$ distinct solutions for $y$ in $S_{0}$.
Plug $ h(x) =  x^{q(q_1+1)} +x^{q_1+1}    +x $ into, and then expanding and simplifying Eq. (\ref{asdaavqefqwdasdasd}), one can obtain
\begin{equation}\label{asdgweasgaegzxfsdfsdfs}
x^q y^{q q_1}+x^{q q_1} y^q+y \left(+x^{q_1}+y^{q_1}+1\right)+x y^{q_1}+y^{q \left(q_1+1\right)} =0 .
\end{equation}
Let $X=x^{q_1}$ and $ Y=y^{q_1}$.
Then Eq. (\ref{asdgweasgaegzxfsdfsdfs}) becomes
\begin{equation}\label{zvavsadvaaf}
	x^q Y^{q }+X^{q } y^q+Xy+Yy+y+x Y+   y^{q}Y^q =0 .
\end{equation}
	Raising both sides of Eq. (\ref{zvavsadvaaf}) to the power of $q$, one can obtain
\begin{equation}\label{agadfadasd}
	x Y+X y+X^{q }y^{q }+Y^{q }y^{q }+y^{q }+x^{q } Y^{q }+   yY=0 .
\end{equation}
By summing Eq. (\ref{zvavsadvaaf})  and Eq. (\ref{agadfadasd}), 
		$y    +y^{q} =0$
is verified.	
Also, $Y    +Y^{q} =0$ is obtained.
For reducing the degree, plugging $Y^{q} =Y$ and $Y=y^{q_1}$ into Eq. (\ref{agadfadasd}) , one can obtain
	\begin{equation}\label{asfasfasdasdq}
		x y^{q_1}+X y+X^{q }y+y+x^{q } y^{q_1}=0 .
	\end{equation}
Clearly $y=0$ is a solution of Eq. (\ref{asdaavqefqwdasdasd}).
Then we assume $y \ne 0$ below.
Easy to obtain
	\begin{equation}\label{}
	( x+x^{q }) y^{q_1-1} +X +X^{q }+1=0.
\end{equation}
Note that $x,X \in S_{\delta}$ and $\tr_{q^2/q}(\delta)\ne 0$ is necessary for $h$ being $ q_1 $-to-$ 1 $.
Then one can obtain
\begin{equation}\label{asdasdasdasdcc}
y^{q_1-1} =1+\frac{1}{\tr_{q^2/q}(\delta)}.
\end{equation}
Since $\gcd \left(  q_1-1, q-1   \right)    =q_1-1$, Eq. (\ref{asdasdasdasdcc}) have $ q_1-1 $ solutions in $\gf_q$ if and only if 
$1+\frac{1}{\tr_{q^2/q}(\delta)}  $ is the $(q_1-1)$-power of some elements in $\gf_q^*$.
\end{proof}
\begin{Exa}
Let $q_1=4, q=2^6$.
According to Proposition \ref{2gouzao1},  $f(x) = \left(      x^{2^6} -x+ \delta  \right)^{5}   +x      $
is a $4$-to-$1$ mapping over $ \gf_{2^{12}} $ if and only if
$\tr_{2^{12}/2^6}(\delta)\ne 0$ and $1+\frac{1}{\tr_{2^{12}/2^6}(\delta)}  $ is a cubic in $\gf_{2^6}^*$.
By a Magma program searching,
$\{
{ \beta^{58}, \beta^{29}, \beta^{36}, \beta^{37}, \beta^9, \beta^{11}, \beta^{42}, \beta^{43}, \beta^{44}, \beta^{45}, \beta^{46},
	\beta^{18}, \beta^{21}, \beta^{50}, \beta^{22}, \beta^{23}, \beta^{53}, \beta^{25}, \beta^{54}, \beta^{27} }
\}$
is the set for $ \tr_{2^{12}/2^6}(\delta) $ exactly satisfying that 
$\tr_{2^{12}/2^6}(\delta)\ne 0$ and $1+\frac{1}{\tr_{2^{12}/2^6}(\delta)}  $ is a cubic in $\gf_q^*$
, where $\beta$ is a primitive element of $\gf_{2^6}$.
And a Magma program verified that $f$ is $4$-to-$1$ exactly when $ \tr_{2^{12}/2^6}(\delta)$ is in this set.
\end{Exa}

\begin{Prop}\label{2gouzao2}
		Let $q$ be a power of  $ 2 $.
	Then,
	$$f(x) =                      \left(      x^{q} +x+ \delta  \right)^{3q}   +x      $$
	is a $2$-to-$1$ mapping over $ \gf_{q^2} $ if and only if $ \tr_{q^2/q}(\delta) \notin \{0,1\}  $.
\end{Prop}
\begin{proof}
	According to Theorem \ref{Zcriterion}, $f$ is $2$-to-$1$
	if and only if 	$ h(x) =   x^{3} + x^{3q}+ x    $ is a $2$-to-$1$ mapping over $S_{\delta} =\left\{z^{q}+z+\delta \  | \  z \in \gf_{q^2}\right\}=\{ z  \in \gf_{q^2}  \   |  \  \tr_{q^2 / q} (z) =\tr_{q^2 / q}(\delta)    \} $.
	It suffices to prove for any $x \in S_\delta$,	
	\begin{equation} \label{agwefwefdzxf}
		h(x+y)=h(x)
	\end{equation}
	has exactly two distinct solutions for $y$ in $S_{0}$.
	Plug $ h(x) =  x^{3} + x^{3q}+ x$ into, and then expanding and simplifying Eq. (\ref{agwefwefdzxf}), one can obtain
	\begin{equation}\label{agwegaegzxfsdfsdfs}
		x^{q} y^{2 q}+x^{2 q} y^{q}+x^{} y^{2 }+x^{2 } y^{}+y^{3 q}+y^{3 }+y=0.
	\end{equation}
Let $X=x^{q}$ and $ Y=y^{q}$.
Then Eq. (\ref{agwegaegzxfsdfsdfs}) becomes
\begin{equation}\label{vascaaasass}
		X Y^{2 }+X^{2}Y+X^{q } Y^{2 q }+X^{2 q } Y^{q }+Y^{3 }+Y^{3 q}+y=0.
\end{equation}
	Raising both sides of Eq. (\ref{vascaaasass}) to the power of $q$, we have
\begin{equation}\label{vasdasdaasd}
		X^{q } Y^{2q }+X^{2q}Y^{q }+X Y^{2 }+X^{2  } Y+Y^{3 q}+Y^{3 }+y^{q }=0.
\end{equation}
By summing Eq. (\ref{vascaaasass})  and Eq. (\ref{vasdasdaasd}), we can verify $y    =y^{q} $ and get $Y    =Y^{q} $.
Plugging $Y^{q} =Y$ and $Y=y^{q}$ into Eq. (\ref{vascaaasass}), one can obtain
\begin{equation}\label{vascaaasassa}
	X y^{2 q }+X^{2}y^{q}+X^{q } y^{2q}+X^{2 q } y^{q}+y=0.
\end{equation}
Clearly $y=0$ is a solution and we assume $y \ne 0$ below.
Since $x,X \in S_{\delta}$, we obtain
\begin{equation}\label{}
	\tr_{q^2/q}(\delta)y+ \tr_{q^2/q}(\delta)^2   + 1=0.
\end{equation}
Clearly $\tr_{q^2/q}(\delta)$ is necessary.
Thus
\begin{equation}\label{}
	y=\frac{ \tr_{q^2/q}(\delta)^2   + 1}{\tr_{q^2/q}(\delta)} \ne 0.
\end{equation}
In summary,  $f$ is $2$-to-$1$ mapping if and only if $ \tr_{q^2/q}(\delta) \notin \{0,1\}  $.
\end{proof}
\begin{Exa}
	Let $q=2^4$. 
	According to Proposition \ref{2gouzao2},  $f(x) = \left(      x^{16} +x+ \delta  \right)^{48}   +x      $
	is a $2$-to-$1$ mapping over $ \gf_{3^6} $ if and only if 
	$\tr_{2^8/2^4}(\delta) \notin \{ 0, 1\} $.
	By a Magma program searching, exactly when $ \tr_{3^6 / 3^3}(\delta) $ in the set
	$\{
	{\beta^1,\beta^2,\beta^3,\beta^4,\beta^5,\beta^6,  \beta^7, \beta^8, \beta^9,\beta^{10}, \beta^{11}, \beta^{12}, \beta^{13}, \beta^{14}   }
	\}, $
	$f$ is $2$-to-$1$, where $\beta$ is a primitive element of $\gf_{3^3}$.
\end{Exa}

\end{document}